\documentclass[useAMS,usenatbib]{mn2e}
\usepackage{mathrsfs}
\usepackage{graphicx}

\def\kms{km s$^{-1}$}   
\def\be{\begin{equation}} \def\ee{\end{equation}} 
\def\deg{$^\circ$} \def\Deg{^\circ}
\def\Msun{M_{\odot \hskip-5.2pt \bullet}}

 \def\htwo{H$_2$}

\def\ROunit{10^{-6}{\rm counts~s^{-1}~arcmin^{-2}}} 
\def\em{{\rm cm^{-6} pc}}

\def\mh{m_{\rm H}}
\def\E{\mathscr{E}}
\def\zsun{Z_\odot}
\def\Zsun{Z_\odot}

\title[Galactic Center Hyper Shell and the North Polar Spur]{Galactic-Center Hyper-Shell Model for the North Polar Spurs}

\author[Y. Sofue et al. ]{ 
Y.Sofue$^{1}$\thanks{E-mail: sofue@ioa.s.u-tokyo.ac.jp},A.Habe$^2$,J.Kataoka$^3$, T.Totani$^4$, Y.Inoue$^5$,S.Nakashima$^5$,H.Matsui$^6$,M.Akita$^3$ 
\\
1. Institute of Astronomy, The University of Tokyo, Mitaka, Tokyo 181-8588, Japan\\
2. Dept. Physics, Hokkaido University, Sapporo 060-0808, Japan\\
3. Research Inst. Science and Engineering, Waseda University, Shinjuku, Tokyo 169-8555, Japan\\
4. Dept. Astronomy, The University of Tokyo, Bunkyo-ku, Tokyo 113-0033, Japan\\
5. ISAS, JAXA, 3-1-1 Yoshinodai, Sagamihara, Kanagawa 252-5210, Japan\\ 
6. National Institute of Technology, Asahikawa College, Asahikawa, Hokkaido 071-8142, Japan
}
\begin{document} 

\date{ }

\pagerange{\pageref{firstpage}--\pageref{lastpage}} \pubyear{20XX}

\maketitle

\label{firstpage}

\begin{abstract}
The bipolar-hyper shell (BHS) model for the North Polar Spurs (NPS-E, -W, and Loop I) and counter southern spurs (SPS-E and -W) is revisited based on numerical hydrodynamical simulations. Propagations of shock waves produced by energetic explosive events in the Galactic Center are examined. Distributions of soft X-ray brightness on the sky at 0.25, 0.7, and 1.5 keV in a $\pm 50\Deg \times \pm 50\Deg$ region around the Galactic Center are modeled by thermal emission from high-temperature plasma in the shock-compressed shell considering shadowing by the interstellar HI and \htwo\ gases. The result is compared with the ROSAT wide field X-ray images in R2, 4 and 6 bands. The NPS and southern spurs are well reproduced by the simulation as shadowed dumbbell-shaped shock waves.  
We discuss the origin and energetics of the event in relation to the starburst and/or AGN activities in the Galactic Center.  
\end{abstract}


\section{Introduction}
 
Nuclear activities in spiral galaxies are evidenced by a variety of ejection phenomena such as jets, rings, lobes, winds and shells of different morphologies and scales from sub parsecs to several tens of kpc, which include those in the Milky Way. (e.g., Oort 1977; Sofue 2000). The discovery of the Fermi Bubble (Su et al. 2010) has stimulated further discussions of the outflow activity in the Galactic Center. The origin and physical properties have been investigated in relation to the associated X-ray features including the North Polar Spur (NPS) (Totani 2006; Fujita et al. 2013; Kataoka et al. 2013, 2015; Ackermann et al. 2014; Mou et al. 2014; Fang and Jiang 2014; Carretti et al. 2013; Crocker et al. 2015; Inoue et al. 2015; Tahara et al. 2015; Sarkar et al. 2015). 

The largest scale Galactic Center (GC) phenomenon so far reported in the Milky Way is the bipolar hyper shells (BHS) with an extent of several kpc and total energy on the order of $10^{55}\sim 10^{56}$ ergs (Sofue 1977, 1984, 1994, 2000; Bland-Hawthorn and Cohen 2003)), which is observed as the North Polar Spur (NPS)  and its counter spurs extending over $\sim 120\Deg$ on the sky in radio continuum (Haslam et al. 1982)  and X-ray emissions (Snowden et al. 1997).  

In our BHS model (Sofue 2000) the giant spurs were interpreted as due to a dumbbell-shaped shock front induced by an explosive event at the Galactic Center $t \sim 15$ Myr ago with total energy of $\E \sim 10^{55}$ ergs. In the present paper, we revisit the BHS model by performing a numerical hydrodynamic simulation of a shock wave driven by high-rate energy injection into the Galactic Center. We simulate soft X-ray distributions and compare them with the ROSAT all-sky maps taking account of shadowing by the interstellar matter.   
 
As to the energy source and its transfer to the kinetic energy of expansion of BHS, we consider the following cases and their combination.
\begin{itemize}
\item SB model: Energy is released by multiple type II supernovae by nuclear starbursts (SB) and is accumulated as kinetic energy to drive a round shock wave or a wide-angle outflow (Sofue 2000). The starburst requires fueling of star forming gas to the central region.
\item AGN model: Energy is released at the black hole in the nucleus (AGN) and transformed to kinetic energy of the surrounding gas to drive a round shock wave or an outflow (Totani 2006;  Mou et al. 2014). 
\end{itemize}
We also categorize the event according to the duration of energy injection.
\begin{itemize}
\item C (continuous injection) type: Energy is continuously released and injected to the Galactic Center. It may occur intermittently by recurrent events. 
\item E (point explosion) type: Energy is released as a single point-like explosive event with a short time scale in the nucleus. If the duration of energy supply in the above C type is sufficiently shorter than the whole event life, it may be regarded as E type. 
\end{itemize}
There may be various hybrid combinations of these models and types.
In the present paper for the BHS, we consider the SB model of C type. This scenario is almost identical to a C type AGN model except for the required gas inflow in SB model.
 
\section{Hydrodynamical Simulation of Bipolar Hyper Shells}

We have performed a numerical hydrodynamical simulation of the BHS by computing the propagation of a shock waves produced by an energy release at the Galactic Center, implicitly considering the SB model with the C type energy injection.

\subsection{Hydrodynamical equations and simulation code}
  
\begin{figure}
\begin{center}
\includegraphics[width=7cm]{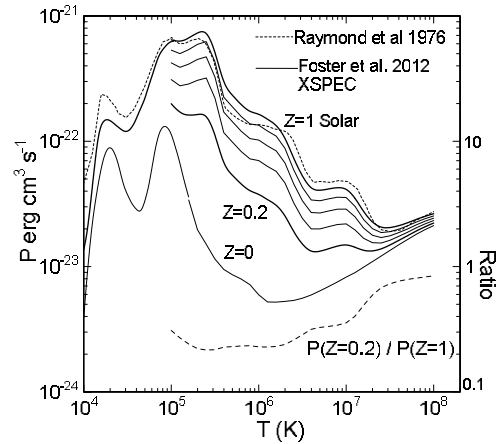}
\end{center}
\caption{Cooling functions $P(T)$ from Raymond et al. (thin dash, 1976) and Foster et al. (2012) generated by XSPEC (Arnauld et al 1997) for $Z=0$ to 1.0. The ratio of the cooling rates for $Z=0.2$ to 1.0 is shown by the thick dashed line.
The solar abundance ($Z=1$) is taken from Anders and Grevesse (1989).
}
\label{cfunc} 
\end{figure}

We used a hydrodynamical code of the flux-splitting method with second-order accuracy both in space and time (van Albada et al. 1982; Mair et al. 1988; Nozawa et al. 2006).  This algorithm is an upwind scheme for the Euler equations and is well suited to solving problems involving a shock. The radiative cooling of gas ($T\ge10^4$ K) is included, assuming the cooling function with the solar abundance (figure \ref{cfunc}).  
 
We assume that gas flow is axisymmetric around the rotation axis of the Galaxy for simplicity.
We solve hydrodynamic equations in the cylindrical coordinate $(r,z)$, assuming that the unperturbed gas in the disk and halo are rotation-supported but the rotation is not important for evolution of bipolar hyper shells.

\def\e{\epsilon}
The hydrodynamic equations of gas are given by
\begin{equation}
\frac{\partial \rho}{\partial t}+\frac{1}{r}\frac{\partial }{\partial r}( r \rho v_r)+\frac{\partial }{\partial z}(\rho v_z)=\dot \rho,
\end{equation}
\begin{equation}
\frac{\partial}{\partial t}( \rho v_r)+\frac{1}{r}\frac{\partial }{\partial r}(r\rho v_r^2)+\frac{\partial}{\partial z}( \rho v_r v_z)=-\frac{\partial p}{\partial r}-\rho\frac{\partial \Phi }{\partial r},
\end{equation}
\begin{equation}
\frac{\partial }{\partial t}(\rho v_z)+\frac{1}{r}\frac{\partial }{\partial r}(r\rho v_rv_z)+\frac{\partial}{\partial z}( \rho  v_z^2)=-\frac{\partial p}{\partial z}-\rho\frac{\partial \Phi }{\partial z},
\end{equation}
\begin{eqnarray} 
\frac{\partial }{\partial t}\left(\rho (\e+\frac{1}{2}v^2)\right) 
+\frac{1}{r}\frac{\partial}{\partial r}\left(r(\frac{\gamma }{\gamma -1}p+ \frac{1}{2}\rho v^2)v_r\right)\nonumber \nonumber \\
+\frac{\partial}{\partial z}\left((\frac{\gamma }{\gamma -1}p
+ \frac{1}{2}\rho v^2)v_z\right) \nonumber  \\
= \dot \E -n^2 P(T)
-\rho \left(v_r \frac{\partial \Phi}{\partial r} + v_z \frac{\partial \Phi}{\partial z}\right).
\end{eqnarray} 
Here, $\rho$ is the gas density, $v_r$ and $v_z$ are  $r$ and $z$ components of the velocity with $v=\sqrt{v_r^2+v_z^2}$, $P(T)$ is the cooling function, $\Phi$ is the gravitational potential of the Galaxy,   $\gamma = 5/3$ is the adiabatic constant for ideal gas, $\e=p/[(\gamma -1)\rho]$ is the specific internal energy,  $\dot \rho$ is the mass injection rate per unit volume,  and $\dot \E$ is the energy injection rate per unit volume. The viscosity, thermal conductivity and the self gravity of the gas are neglected. 
The cooling  function $P(T)$ was taken from Raymond et al. (1976) for solar abundance as shown in figure \ref{cfunc}. Note that the cooling function is almost identical to that by Foster et al. (2012) in the temperature range under consideration.
 
\subsection{Circumstantial and initial conditions}

The gravitational potential $\Phi$ is assumed to be given by the Miyamoto and Nagai (1975) model,  
\begin{equation}
\Phi(r, z)=-\frac{GM}{\sqrt{r^2+({a+\sqrt{z^2+b^2}})^2}}, 
\end{equation}
where $G$ is the gravitational constant, $M=10^{11}M_{\odot}$, $a=20$ kpc, and $b=$1 kpc.

The initial gas density was assumed to have the distribution represented by
\begin{equation}
\rho(r, z)= \rho_1 \exp(-(z/z_1)^2))+\rho_2(-(z/z_2))+\rho_3.
\label{initialgas}
\end{equation}
Here $\rho$ is the density, $z$ is the height from the galactic plane, $z_i$ is the scale thickness of the disk and halo, and $r$ is the distance from the rotation axis. Suffices 1, 2 and 3 denote quantities representing the disk, halo and intergalactic gas, respectively, with $\rho_1=3\times 10^{-24}$g cm$^{-3}$, $\rho_2=3\times 10^{-26}$g cm$^{-3}$, and $z_1$ and $z_2$ are  0.1 and 1 kpc. The last term represents the intergalactic gas of density on the order of $10^{-29}$g cm$^{-3}$.  
 The initial temperatures in the disk and halo gases were taken to be $10^4$ K and $2\times 10^6$ K, respectively.
The thus settled gas distribution is in between the plane-parallel (Yao et al. 2009; Sakai et al. 2014) and spherical (Miller and Bregmann 2013) models for the hot halo gas distribution as inferred from X-ray observations.

The numerical simulations were performed for the two cases, one for E type injection with a point-like explosion at the nucleus, and the other for C type with continuous energy injection into the Galactic Center. In both cases total energy of $\E=4\times 10^{56}$ ergs was given as thermal energy into a sphere of 50 pc radius. In the C type injection, the energy was supplied to the volume continuously at a rate of $d\E/dt=4.0\times 10^{55}$ ergs Myr$^{-1}$, and total energy of $\E=4\times 10^{56}$ ergs in 10 Myr, where we also assumed mass supply of at a rate of $dM/dt=1\Msun{\rm yr}^{-1}$.

The initial gas was left free in the gravitational potential with the distribution mimicking multiple layered disks composed of a low-temperature disk and high-temperature halo by equation \ref{initialgas}. The gas is at rest initially in the potential, but moves slowly toward hydrostatic distribution. Since the motion is slow, the ambient gas appears almost motionless in contrast to the rapid shock wave propagation. The gravitational and hydrodynamical accelerations are included in the equation of motion.
 
\subsection{Metallicity and cooling}

In the present simulation, the solar abundance was assumed not only because the code was written for a fixed solar abundance, but also because the metallicity in the Galactic Center and halo is not well modeled. The gas in the BHS is a mixture of different components with different abundances from extremely high to primordially low. Namely, the BHS includes the ejected/snowplowed gas from the galactic disc and the Galactic Center with high star formation rate and therefore high metallicity, $> Z_\odot$ (Uchiyama et al. 2013; Najarro et al. 2009), and the halo gas for which some authors prefer high metallicity of $ Z=\Zsun$ (Sakai et al. 2014) but the others prefer lower value at $Z=0.2 \Zsun$ (Miler et al. 2014; Kataoka et al. 2015). Also, intergalactic gas supposed to have lower metallicity may be included.

So, we here assume solar abundance $Z=1\Zsun$ as an average during the whole life of the BHS evolution. The metallicity influences the simulation through cooling rate. However, it is shown in section 4, that the cooling does not affect the BHS dynamics and evolution significantly.
Note also that we adopted the observed metallicity, $Z=0.2\Zsun$, when calculating the 'present' brightness of the NPS for comparison with the observations.
 
\subsection{Results for C (BS) vs E (AGN) type energy injections}

\begin{figure} 
\begin{center} 
\includegraphics[width=8cm]{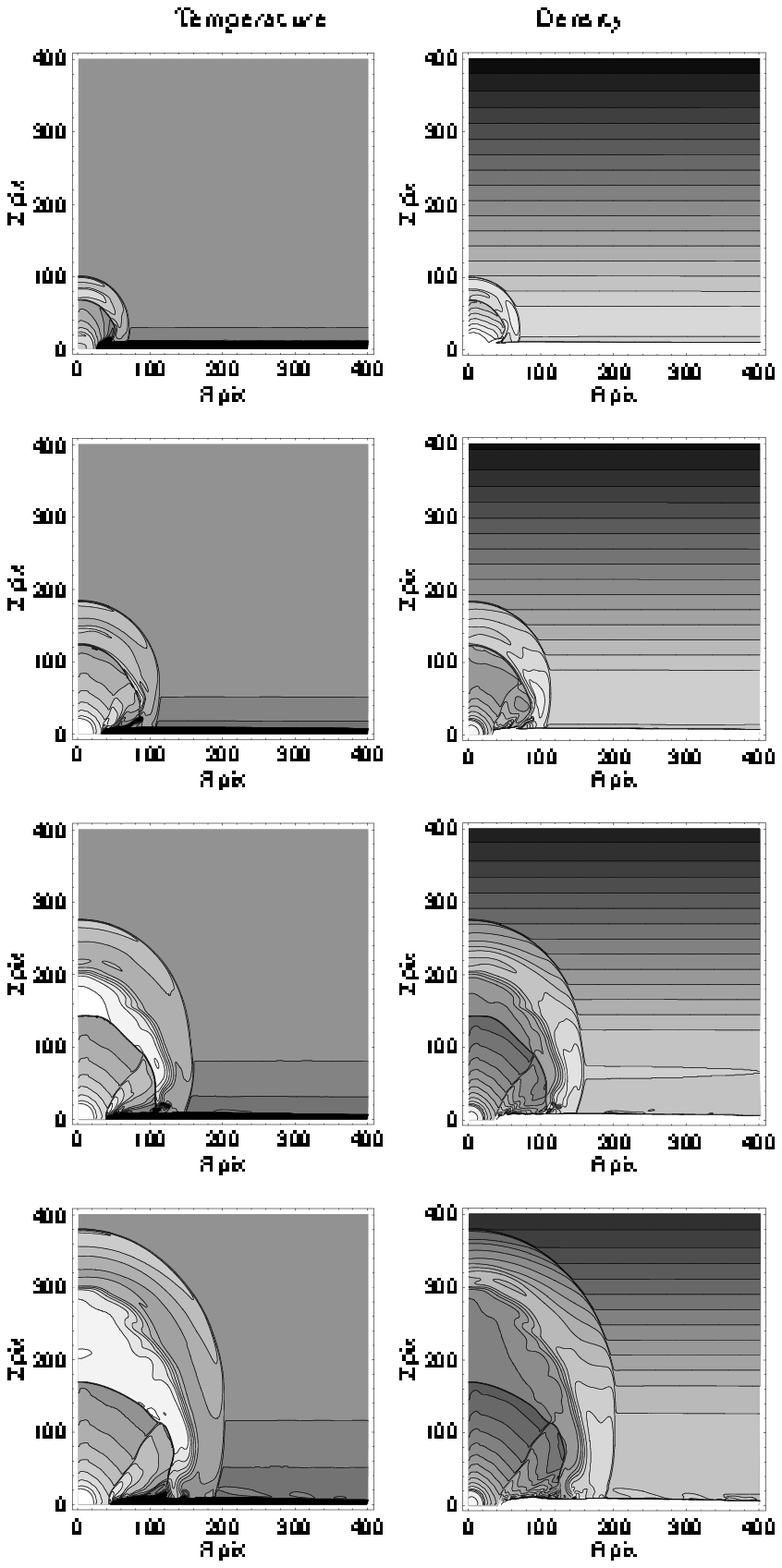}\\
\end{center}
\caption{Hydrodynamical simulation of type C (continuous) injection model for density  and temperature  distributions from $t=2$ to 7 Myr. Density contours are drawn at log $\rho$ (H cm$^{-3})=-4$ (black) to $-1$ (white) at equal dex interval $\Delta$ log $\rho=0.2$, and temperature from log $T$ (K)=5 to 8 at dex interval 0.2. One pixel corresponds to 20 pc and the frame covers 8 kpc by 8 kpc region with the Galactic Center at the lower left corner and the galactic plane horizontal.} 
\label{habe1} 
\end{figure}

\begin{figure*} 
\begin{center} 
\includegraphics[width=14cm]{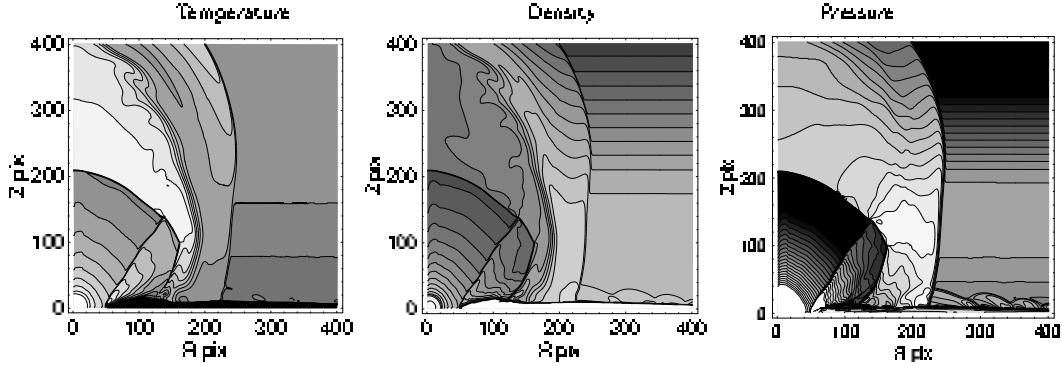} 
\end{center}
\caption{Same as figure \ref{habe1} at 10 Myr for the temperature, density and the pressure. The pressu contours are at every 0.1 dex from $10^{-13}$ to $10^{-11}$ ergs cm$^{-3}$. This result is used for the analysis and comparison with the observations of the NPS. }  
\label{habe10} 
\end{figure*}

\begin{figure} 
\begin{center}
\hskip -8mm \includegraphics[width=8cm]{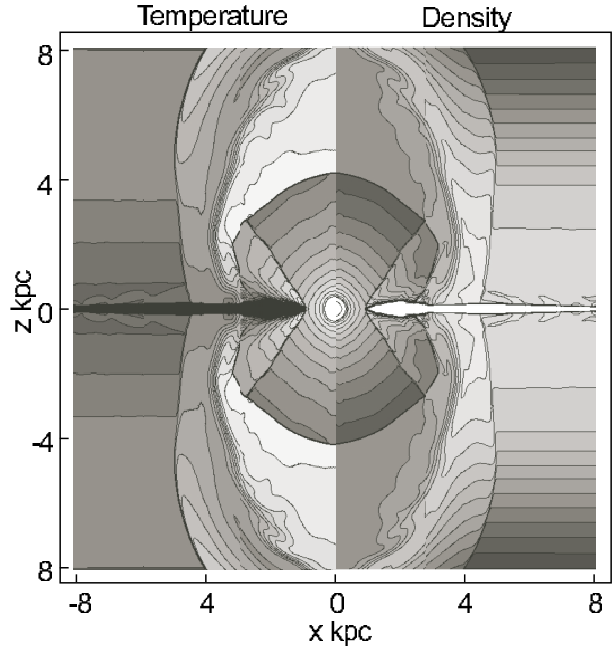} 
\end{center}\caption{Same as figure \ref{habe1} at 10 Myr for the density and temperature for convenience of comparison and correspondence. This figure reveals the bipolar hyper shell (BHS) structure of the shock front propagating in the galactic disk and halo. Also interesting feature is seen as the inner bipolar cone feature produced by a high-velocity outflow as a bow shock against the galactic disk. } 
\label{habe2} 
\end{figure}

The calculated result for C (continuous; BS like) type injection is shown in figures \ref{habe1}, \ref{habe10} and \ref{habe2}. 
The shock front expands spherically in the initial $\sim 1$ Myr, and, then, it is elongated into the halo, composing dumbbell shaped symmetric bubbles with respect to the galactic disk, which we call the BHS. Detailed description of the structure and evolution is given in the next subsection. 

In this paper, we describe the result of C type in detail, which was found to fit the NPS morphology better than the E type result. However, we do not intend to deny the E type event, which might be simpler to explain the shell phenomena without worrying about multiple, complicated phenomena recurrently occurring in the Galactic Center.
 
\begin{figure} 
\begin{center}
Temperature ~~~~~~~~~~~~~~~~~~~~~ Density\\
\hskip-2mm\includegraphics[width=4.2cm]{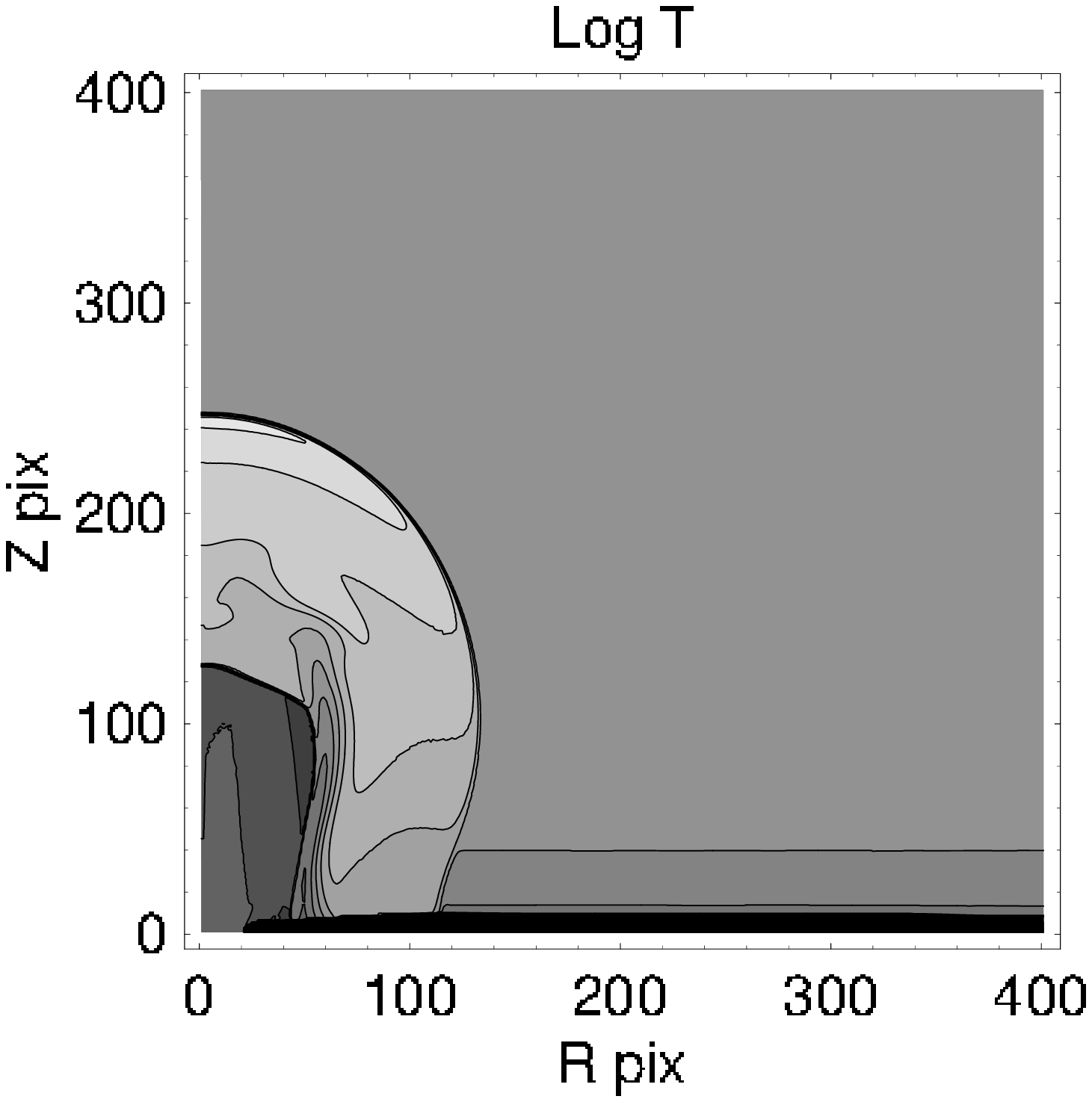} 
\includegraphics[width=4.2cm]{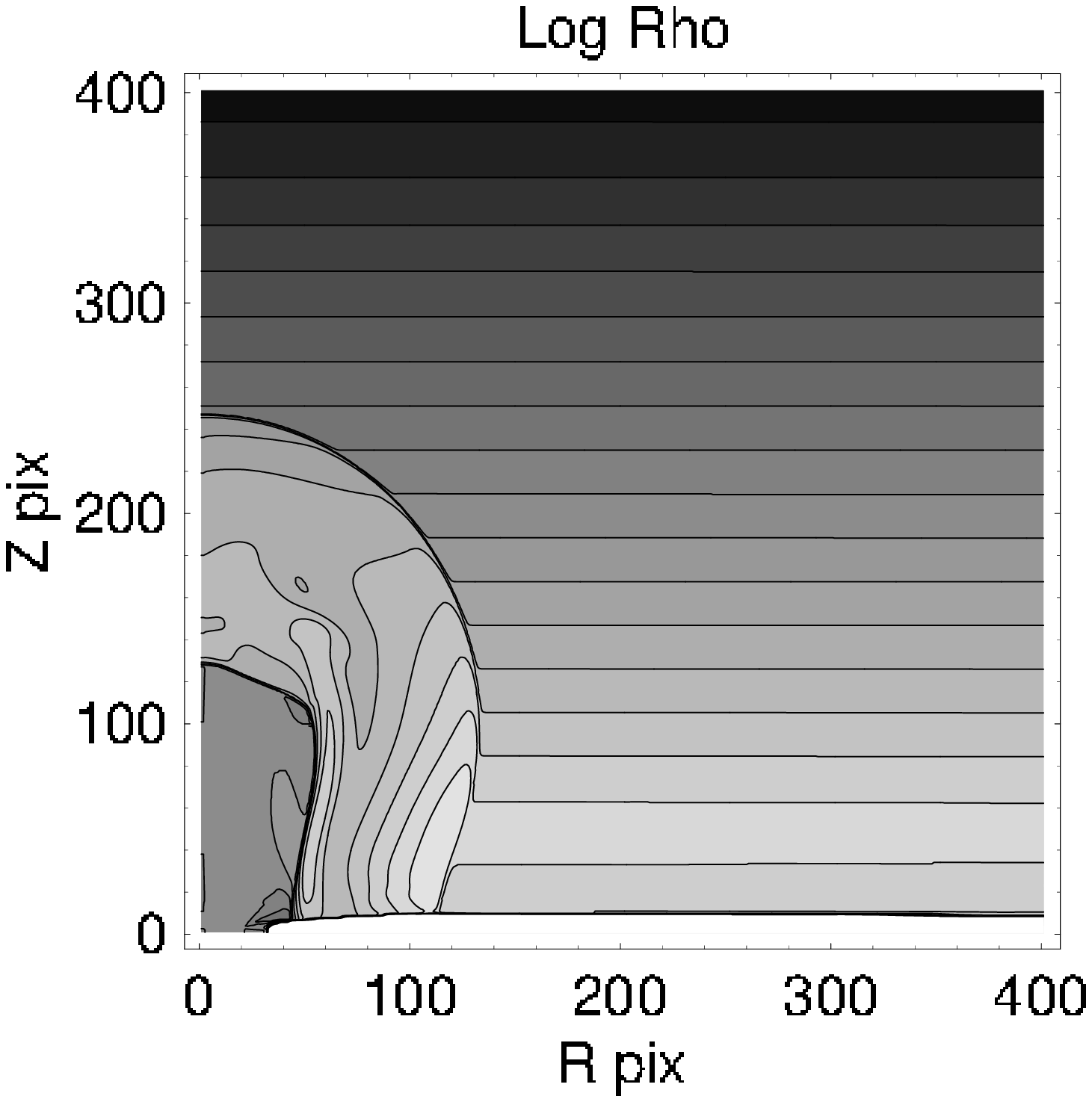} 
\end{center}

\caption{Same as figure \ref{habe1} at 3 Myr for E type energy injection. The BHS resembles that for C type, while the central region is smoother and the cone feature does not appear. } 
\label{habePE} 
\end{figure}

For comparison, we show the result for E (point explosion; AGN like)-type injection in figure \ref{habePE} for $t=3$ Myr. The global structure and evolution of the produced shock wave shell is similar to that for C-type. However, the general shock wave structure is much simpler. The highest density appears at lower latitudes, whereas the highest temperature was attained in the polar region in the front shock facing the halo gas. 
The central region has much smoother structure compared to that for C type, having a simple cavity of low density and low temperature. Since no wind blows, no bow-shock or a cone structure is produced near the disk.

\subsection{BHS by C-type injection for SB model}

\begin{figure} 
\begin{center}
\includegraphics[width=7cm]{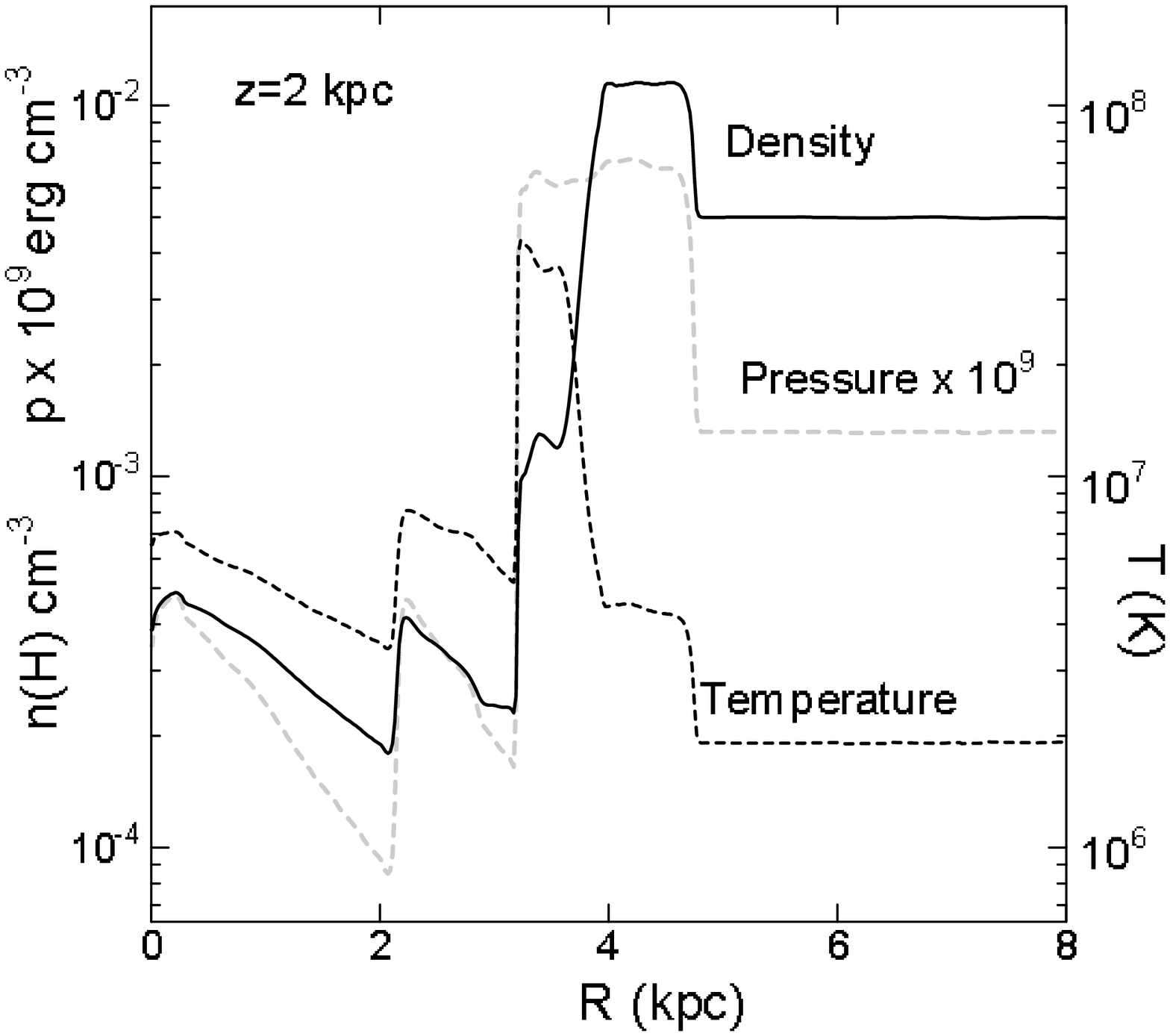}
\includegraphics[width=7cm]{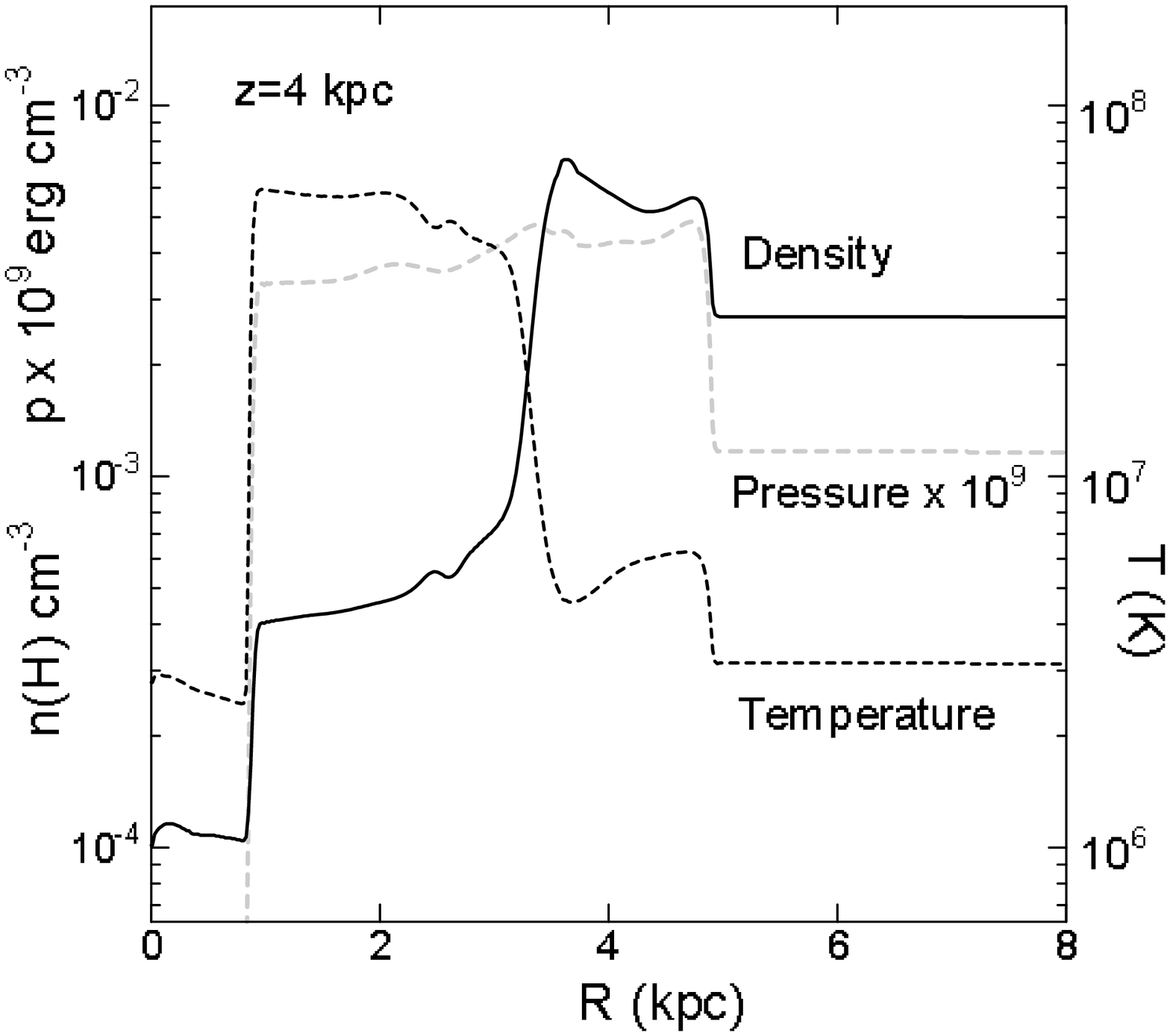}
\end{center}
\caption{Density (full line), temperature (dashed) and pressure (gray dash) distributions at constant height $z=2$ and 4 kpc at $t=10$ Myr. } 
\label{habecross} 
\end{figure}

\begin{figure} 
\begin{center}
\includegraphics[width=7cm]{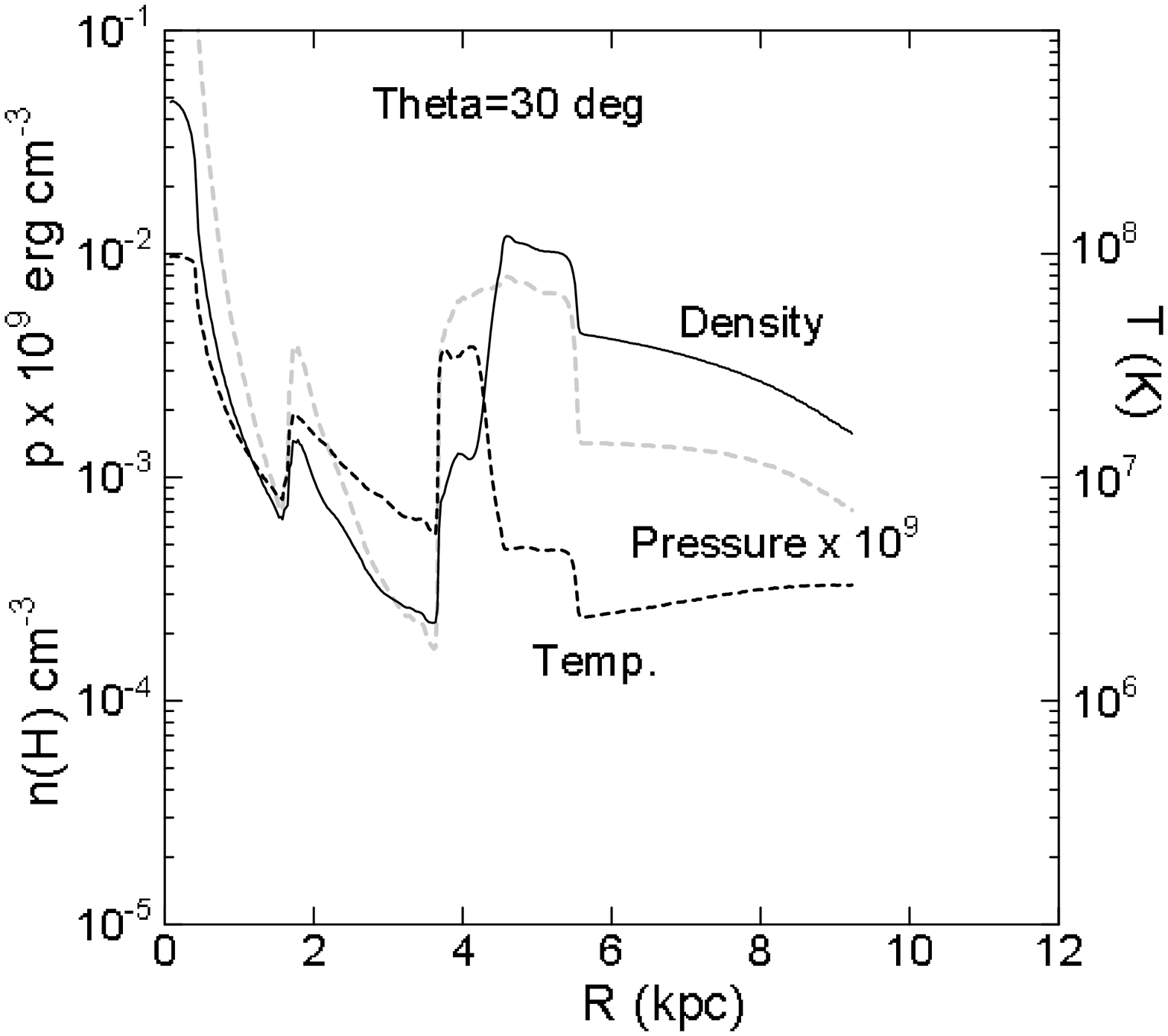} 
\includegraphics[width=7cm]{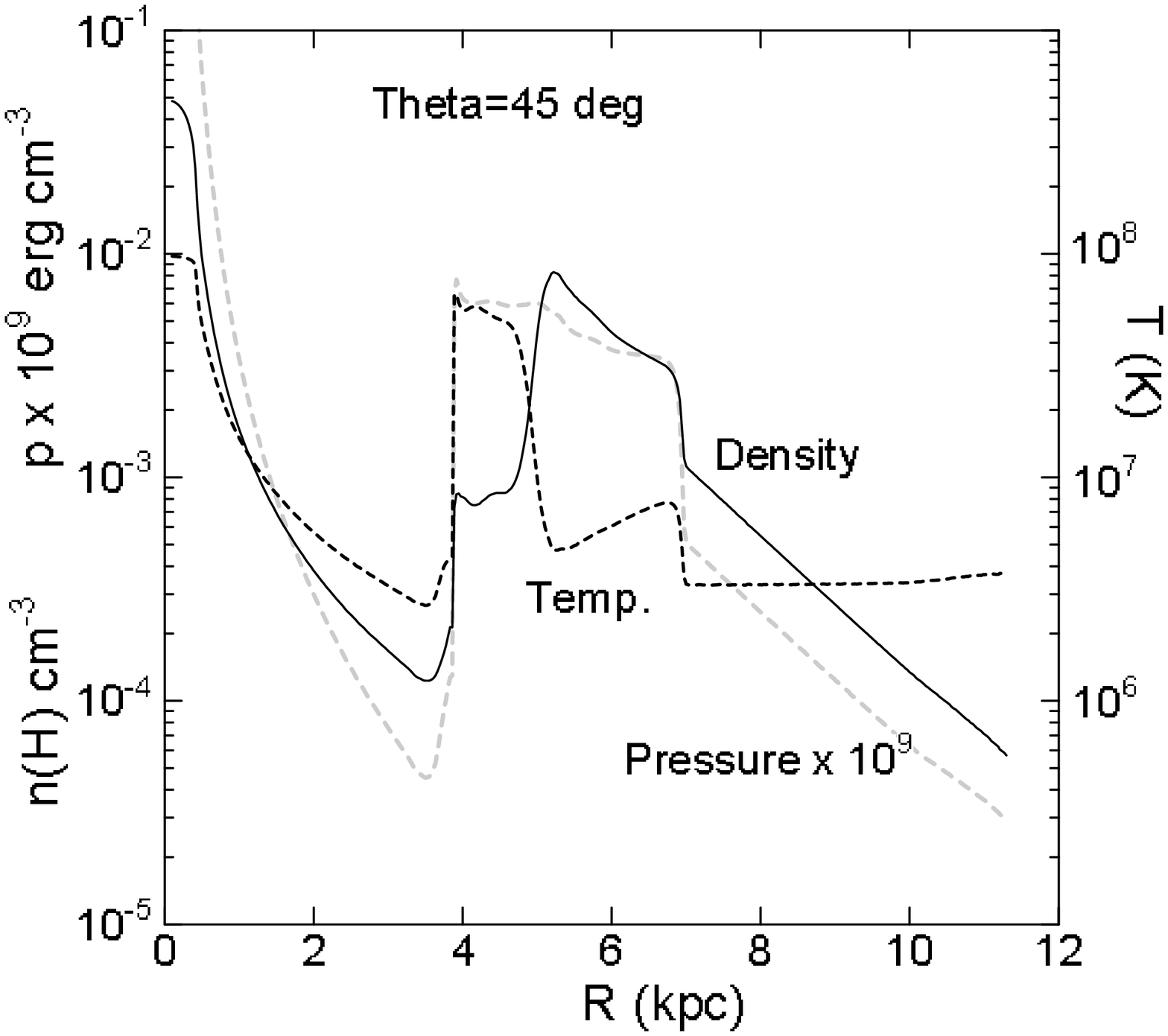} 
\end{center}
\caption{Radial distributions of density (full), temperature (dash), and pressure (gray dash) at altitude angle 30\deg and 45\deg at $t=10$ Myr.
} 
\label{habecross2} 
\end{figure}

The simulated result for the C-type BHS evolution corresponding to the SB model  is shown in figures \ref{habe1} to \ref{habe2}. 
Figure \ref{habecross} shows cross sections along  constant heights at $z=2$ and 4 kpc at $t=10$ Myr, and figure \ref{habecross2} shows radial cross sections at altitude angle $\Theta=30\Deg$ and $45\Deg$ from the galactic plane. 

 The central disk within a few hundred pc is disrupted to yield a hole inside a ring around the nucleus. Due to the continuous injection of mass, there appears an outward flow. When the outflow encounters the wall of the hole, it produces a shock-compressed ring, and a bow shock is produced, which extends into the halo as a cone-shaped, high-pressure discontinuity. By the constantly injected energy, the central region is kept at high temperature and high density. We will discuss the central reion in more detail in section 4.

When the outflow encounters the halo gas, it produces round-shaped shock wave. The shock front velocity is decelerated near the galactic plane by the interaction with the dense ambient gas. 
The shock front expands spherically in the initial $\sim 1$ Myr, and is elongated into the halo as the time elapses. 
In the halo, the expanding velocity increases due to energy injection at the center and as well as by the jet-acceleration mechanism of Sakashita (1971) in exponentially decreasing gas density.

When BHS reaches high-altitude at $z\sim 6$ kpc and encounters the intergalactic gas, the front suffers from decelerating force. However, due to the continous supply of gas and energy from inside, the front still continues to expand at increasing velocity.
 
Figure \ref{habe2} shows enlargement of the result at 10 Myr with the density and temperature in a mirror plot for convenience of comparison and recognizing the mutual correspondence. The figure reveals more clearly the bipolar hyper shell (BHS) structure of the outer shock front propagating in the halo. Also impressive in this figure is the inner bipolar conical horn feature, which is a bow shock against the galactic disk produced by a high-velocity wind from the center (see the discussion section).

As shown in figures \ref{habecross} and \ref{habecross2} the densest shell follows the shock front, while the highest temperature is attained inside the shell, facing the cavity. The dense shell, where the emission measure is largest, has temperature around $ \sim 0.3$ keV ($3.5\times 10^6$ K), in agreement with the observed temperature in the NPS (Kataoka et al. 2015). On the other hand, the high temperature inner shell facing cavity, where $T\sim 4$ keV ($\sim 5\times 10^7$ K), has density one to two orders of magnitude less than that in the dense shell. Hence, the emission measure in the high temperature cavity is much lower than in the dense shell, so the X-ray emission is much weaker compared to the shell.

\section{Soft X-ray Sky}
 
\subsection{Emission Measure}

Using the calculated density distribution by the BHS model, we calculated the emission measure along the line of sight $s$, 
\begin{equation}
EM=\int (\rho/\mh)^2 ds.
\label{eqEM}
\end{equation} 
During the integration, we avoided the contribution from the ambient disk and halo gases by setting a minimum temperature of gas at $T_{\rm min}=3\times 10^6$, which is the halo gas temperature as represented by flat temperature beyond 5 kpc in figure \ref{habecross}. By this condition, the lower temperature disk is also avoided from integration. This limitation does not cause significant under estimation of $EM$ in the NPS, because the temperature inside the BHS is mostly higher than $3\times 10^6$ K (figure \ref{habecross}).

\begin{figure} 
\begin{center}
\includegraphics[width=6.5cm]{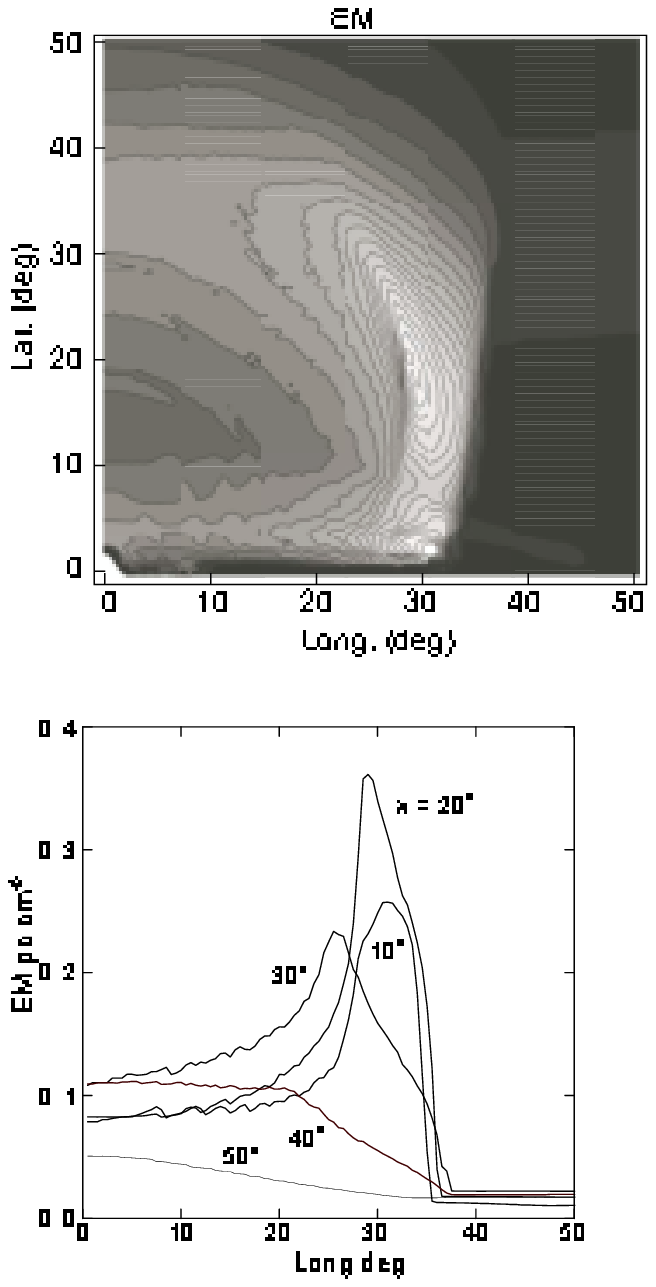} 
\end{center}
\caption{[Top] Dstribution of the emission measure $EM$ on the sky at $t=10$  Myr. 2 pixles corresponds to 1 degree. [Bottom] Cross sections along constant latitudes at $b=0\Deg$ to $50\Deg$.
}
\label{figEM} 

\begin{center} 
\includegraphics[width=7cm]{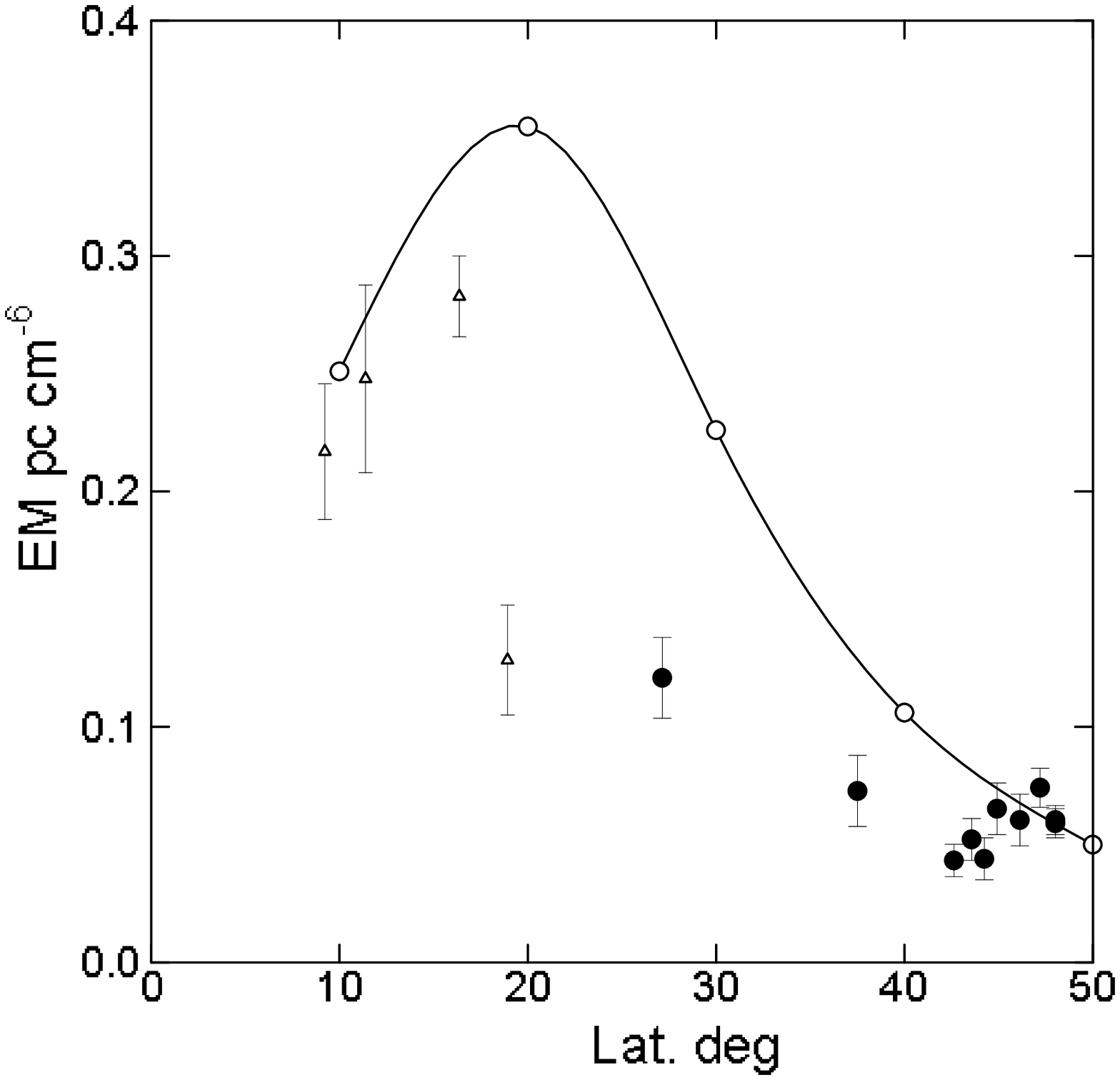} 
\end{center}
\caption{[ Variation of $EM$ with $b$ along the BHS ridge (full line), compared with measured $EM$ by Kataoka et al. (2015) along the NPS (big dots). Small triangles are southern Claw and NE clump inside the Fermi Bubble, and hence not to be compared here. 
} 
\label{figEMvsSuzaku} 
\end{figure}

A result for $t=10$ Myr is shown in figure \ref{figEM} by a contour map in the $50\Deg\times 50\Deg$ sky, as well as by cross sections along constant latitudes, $b=20\Deg$ and $30\Deg$. The peak values across the BHS are obtained to be $EM\simeq 0.1 - 0.37 \em$ at $b=10\Deg$ and $40\Deg$.
 
Figure \ref{figEMvsSuzaku} shows the simulated $EM$ plotted against latitude. Plotted together are the $EM$ values determined by  {\it Suzaku} observations along the NPS at $b\sim 10-50\Deg$ (Kataoka et al. 2015). 
The calculated values are greater than the observed values, which are $EM\sim 0.06-0.3$, by a factor of $\sim 1.5$. Similar values, $\sim 0.1~ \em$, have been obtained by Willingale et al. (2003) based on a local shell origin of NPS.  

\subsection{Emission}

\begin{figure}  
\begin{center} 
(a)\includegraphics[width=7cm]{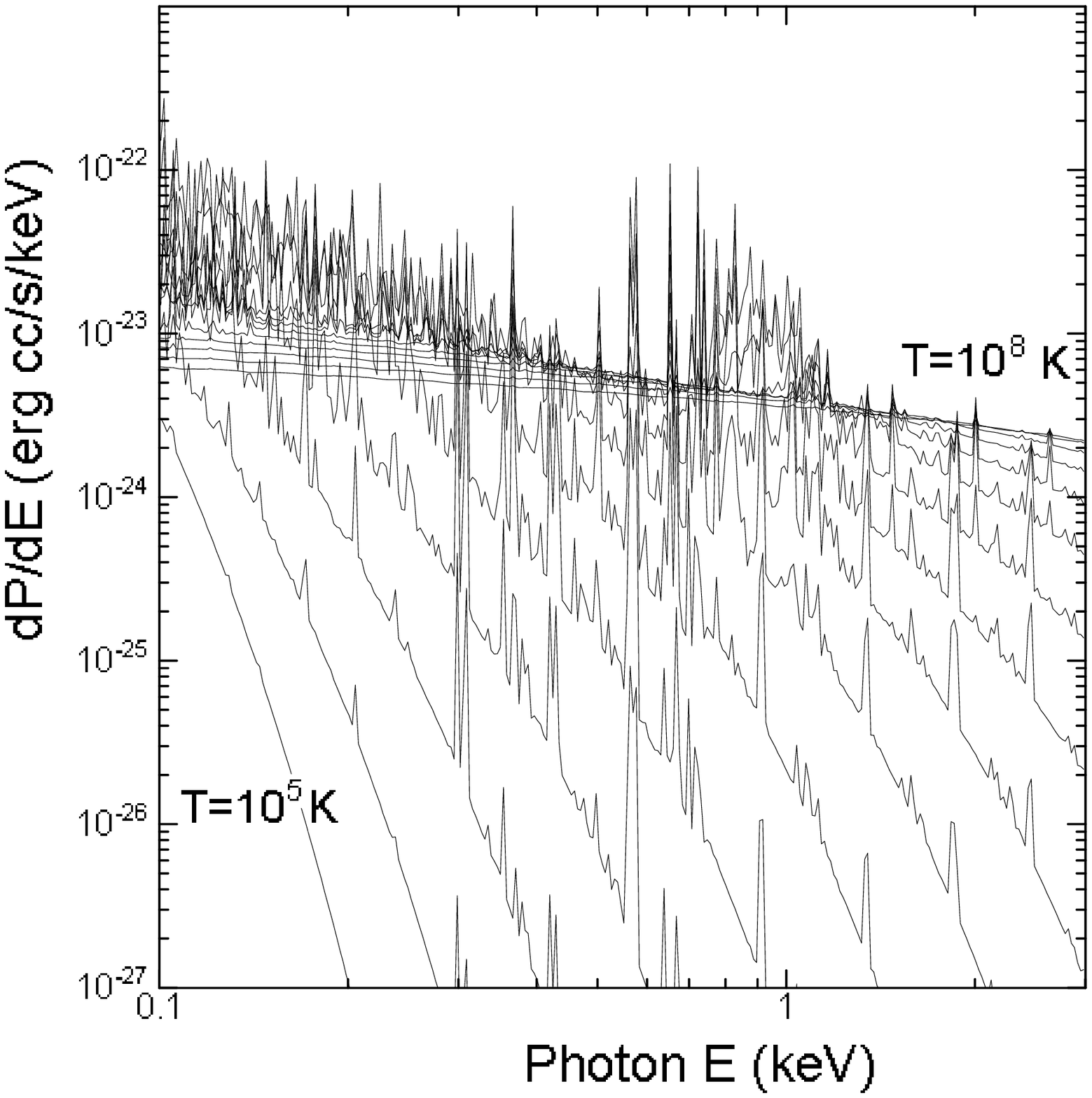} 
(b)\includegraphics[width=7cm]{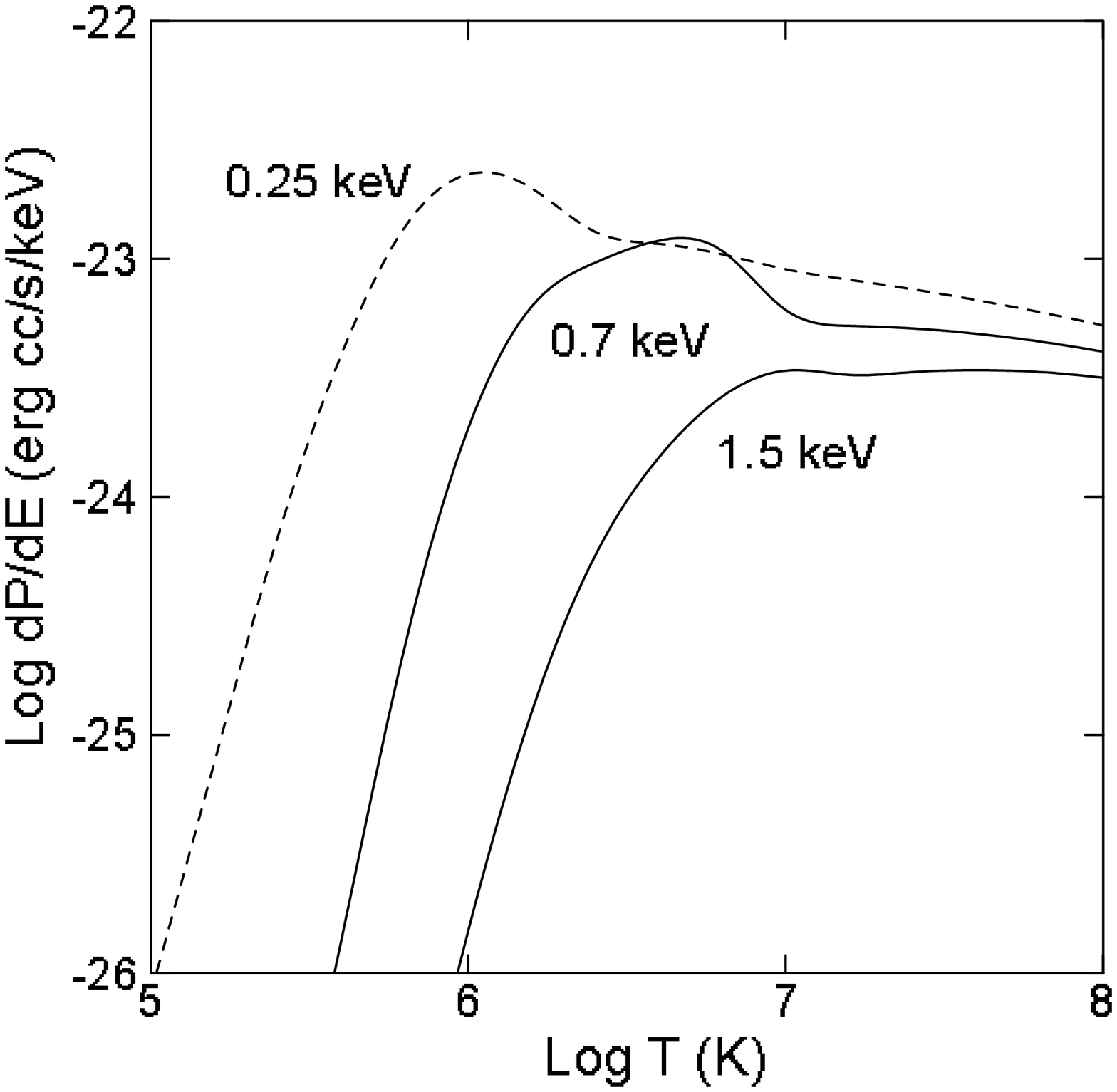}  
\end{center}
\caption{(a) Spectral emissivity for $Z=0.2 Z_\odot$ against emitted photon energy for different plasma temperatures at every 0.2 dex temperature from $T=10^5$ to $10^8$ K as calculated by Xspec (Arnauld 1996; Foster et al. 2012). (b) Spectral emissivity against temperature for fixed emitted photon emerges at $E=0.25$, 0.7 and 1.5 keV averaged by Gaussian band response with full width of $\Delta E=0.5 E$ as calculated by using (a). } 
\label{cfuncXspec} 
\end{figure}

We calculated the emissivity of X-rays using the computed density and temperature, assuming a metallicity of $Z=0.2Z_\odot$ and cooling functions given by Foster et al. (2012) as shown in figure \ref{cfuncXspec} obtained by using XSPEC (Arnauld et al. 1996).  The surface brightness on the sky at photon energy $E$ was calculated by integrating the emissivity along the line sight,
\begin{equation}
B={1 \over 4 \pi} \int {dP(E) \over dE} n_e^2  ds.
\label{B}
\end{equation} 
Here, $dP(E)/dE$ is the spectral emissivity per unit volume and unit density per unit photon energy $E$,  and $n_e$ is the electron density assumed to be equal to the ion density $n_e\simeq \rho/m_{\rm H}$.  The brightness $B$ is expressed in Jy str$^{-1}$. 

The spectral emissivity was averaged by $E$, assuming a Gaussian response function with a full width of half maximum $\Delta E=0.5E$ around the center energy $E$. Figure \ref{cfuncXspec} shows the thus obtained spectral emissivity as functions of the plasma temperature for fixed photon energies at $E=0.25,$ 0.7 and 1.5 keV, approximately representing the ROSAT R2, 4 and 7 bands. 

\subsection{Absorption and shadowing}

\begin{figure} 
\begin{center}
\includegraphics[width=7cm]{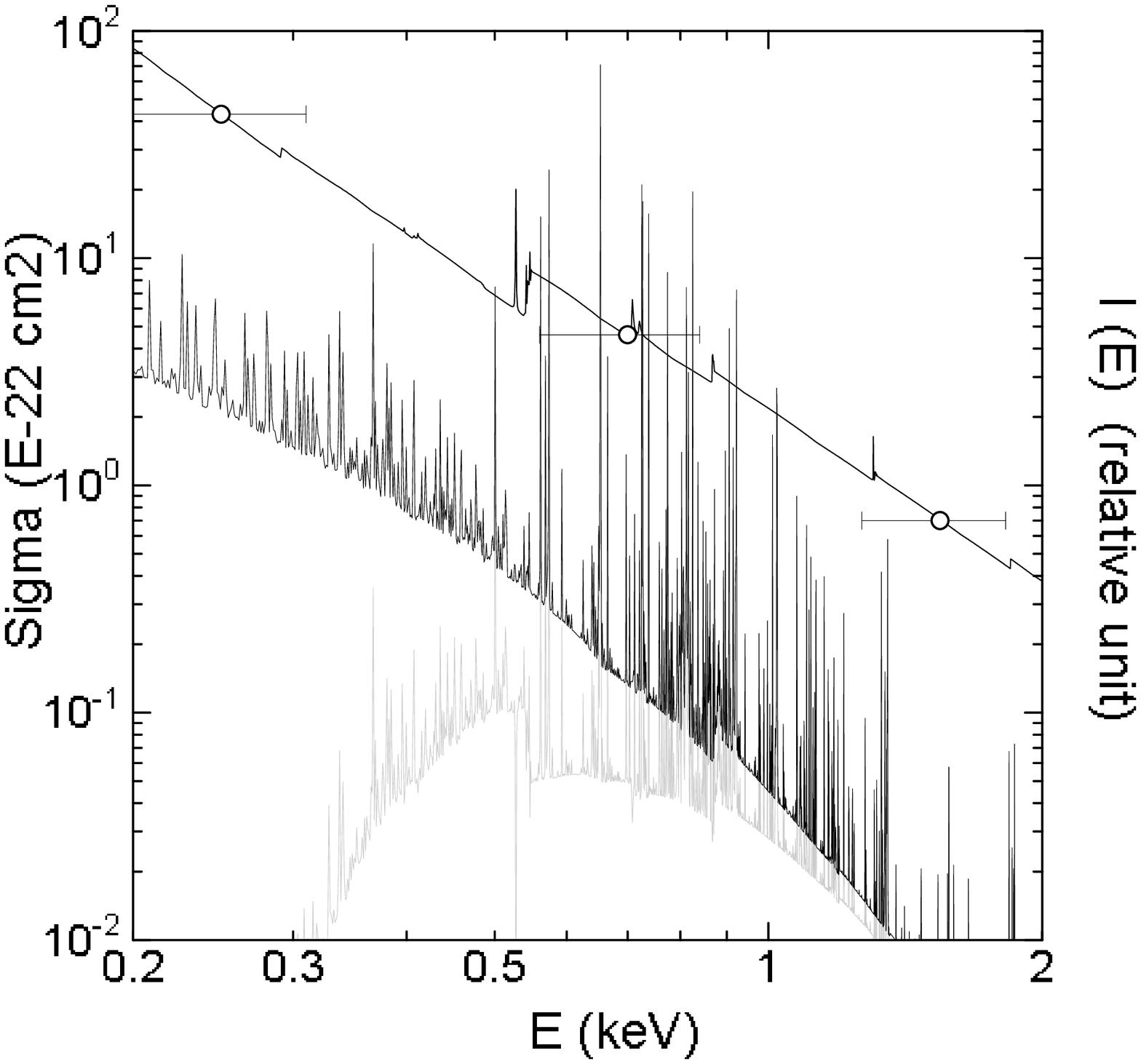} 
\end{center}
\caption{Absorption cross section for $Z=1 Z_\odot$ (upper line), spectral emissivity $dP(E)/dE$ (second line) for $E=0.3$ keV, and $dP(E)/dE ~\exp (-\sigma/\sigma_4)$ (gray line) for R4 band as obtained by ISMabs (Gatuzz et al. 2015) in XSPEC (Arnauld et al. 1997), demonstrating that the transmitted X-ray emission has nearly flat spectrum through a gas around $\tau \sim 1$.
Approximate R2, 4 and 7 band energy and band widths are marked by circles. } 
\label{sigma}  

\begin{center}
\includegraphics[width=7cm]{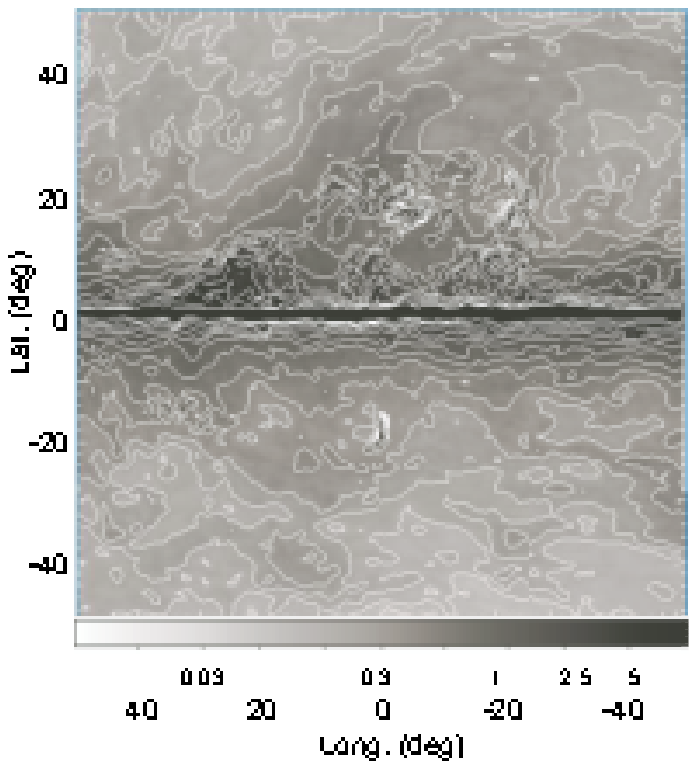} 
\end{center}
\caption{Optical depth for X-rays in R4 band, which is equal to the column density divided by the critical column density,  $\tau_4=N({\rm H})/N_4$, where $N({\rm H})=N({\rm HI})+2N({\rm H}_2)$. Contours are drawn at logarithmic interval of 0.05 dex from $\tau_4=0$ to 10. Optical depth for other bands are obtained by $\tau_i=\tau_4 N_4/N_i$.} 
\label{tau4} 
\end{figure}

The optical depth of X-ray emission is defined by $\tau=N \sigma_i$, where $\sigma_i$ is the absorption cross section for the $i$-th band. X-ray absorption coefficient by metals in the interstellar gas with solar abundance was calculated by using 'ISMabs' (Gatuzz et al. 2015) implemented in XSPEC as shown in figure \ref{sigma}, and the adopted values are listed in table \ref{tabband}.

The H atom column density $N$ was calculated using observed integrated intensities of the HI and CO ($J=1-0$) line emissions. We used the Argentine-Leiden-Bonn All-Sky HI Survey (Kalbelra et al. 2005) and the Columbia Galactic Plane CO Survey (Dame et al. 2002). The total column density $N({\rm H})$ of hydrogen atoms is obtained by
\be
N({\rm H})= C_{\rm HI} I_{\rm HI} +2C_{\rm H_2} I_{\rm CO},
\ee
where $I_{\rm HI}$ and $I_{\rm CO}$ are integrated HI and CO line intensities. 

The HI conversion factor was taken to be 
$C_{\rm HI}=1.82\times 10^{18}{\rm cm^{-2}} /{\rm K~km~s^{-1}}$. 
The CO-to-H$_2$ conversion factor was assumed to be 
$C_{\rm H_2}=2.0 \times 10^{20} {\rm H_2 cm^{-2}} /{\rm K~km~s^{-1}}$ for the local interstellar gas having the solar abundance (e.g. Bollato et al. 2013).
Figure \ref{tau4} shows a distribution map of the optical depth $\tau_4=N/N_4$ for R4 band X-rays, which is equivalent to the distribution of the total H atom column density, except that the values are divided by $N_4$. Optical depth maps for other bands are obtained by multiplying $\tau_i=\tau_4 N_4/N_i$.  

The absorption was calculated for fixed photon energy in each band, and hence, the model intensity distribution represents a monochromatic view of shadowed sky. In observations, however, the photon spectra are effectively hardened by the absorption within each band. However, the emissivity itself is a decreasing function of $E$. Figure \ref{sigma} shows the absorption cross section and spectral emissivity, and the transmitted emission through gas with optical thickness unity , $\tau=1$. The figure demonstrates that the absorption and emission dependencies with photon energy act to cancel each other, so that the transmitted emission has nearly a flat spectrum. Thus we may conclude that the monochromatic treatment is a good approximation to the band-averaged surface brightness, sufficient for the present analysis.

\begin{table}
\caption{Representative energies for R2, 4 and 7 bands used for simulation, and band-averaged  absorption cross sections calculated using 'ISMabs' (Gatuzz et al. 2015) implemented in XSPEC (Arnauld et al. 1997).}
\begin{center}
\begin{tabular}{lllll}  
\hline
Band&  $E$ & $\sigma_i$ &$N_i=1/\sigma_i$\\
& keV   &  $10^{-22}$ cm$^2$ &$10^{20}$ cm$^{-2}$ \\
\hline  
R2 & 0.25 & 43.& 2.3 \\ 
R4 &0.7 &5.0 &  20.0 \\ 
R7 &1.5 &0.7&  142.  \\
 
\hline
\end{tabular}
\end{center}
\label{tabband}
\end{table} 

\subsection{Brightness}

\begin{figure*} 
\begin{center}  
(a)~~~~~~~~~~~~~~~~~~~~~~~~~~~~~~~~~~~~~~~(b)~~~~~~~~~~~~~~~~~~~~~~~~~~~~~~~~~~~~~~~~(c)\\
\includegraphics[height=6cm]{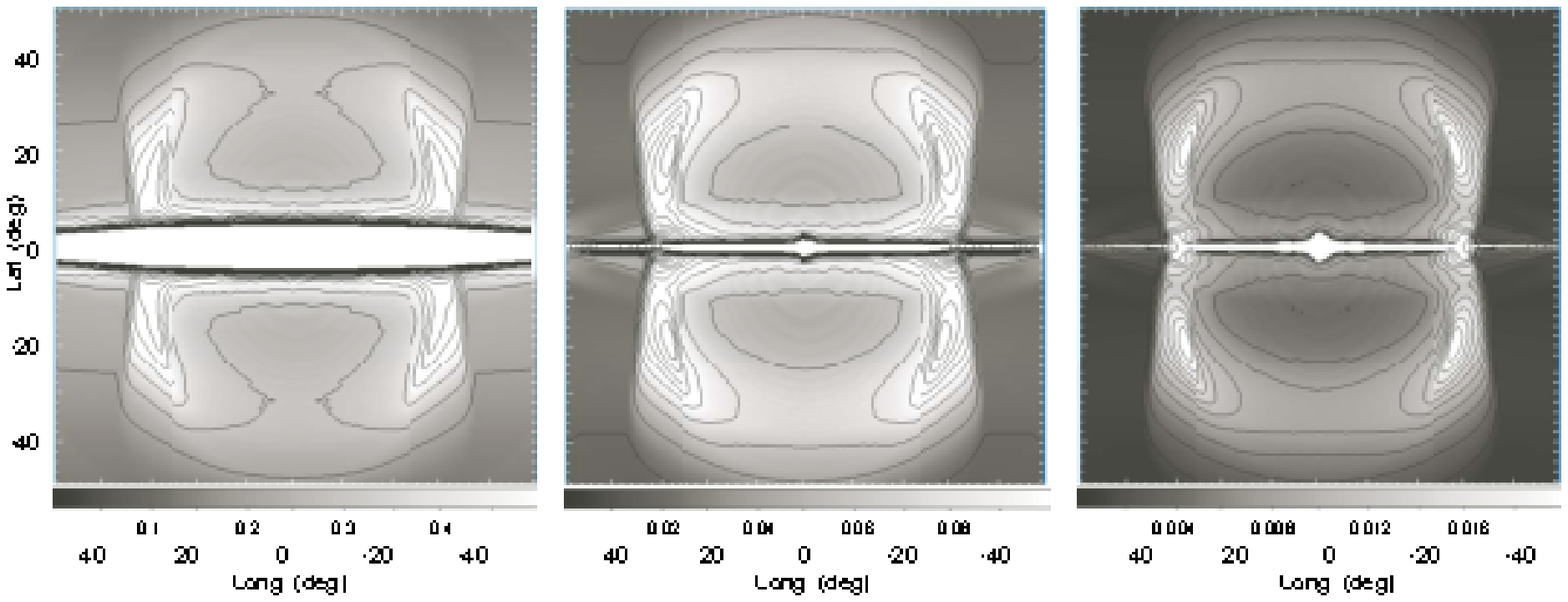}  
\end{center}
\caption{(a) Simulated brightness distribution at 0.25 keV ($\sim$R2) at $t=10$ Myr in the central $\pm 50\Deg \times \pm 50\Deg$.  Contours are at
0, 0.1, 0.2, ...., 1.9, 2.0 0.2 Jy str$^{-1}$. (b) Same, but 0.7 keV ($\sim $R4) with contours at 0, 0.01, 0.02, ..., 0.19, 0.2 Jy str$^{-1}$. (c) Same, but 1.5 keV ($\sim$R7) with contours at 0, 0.002, 0.004, ...., 0.018, 0.02 Jy str$^{-1}$.
} 
\label{xsky1}  

\begin{center}
(a) ~~~~~~~~~~~~~~~~~~~~~~~~~~~~~~~~~~~~~~~ (b) ~~~~~~~~~~~~~~~~~~~~~~~~~~~~~~~~~~~~~~~ (c)\\
\includegraphics[height=6cm]{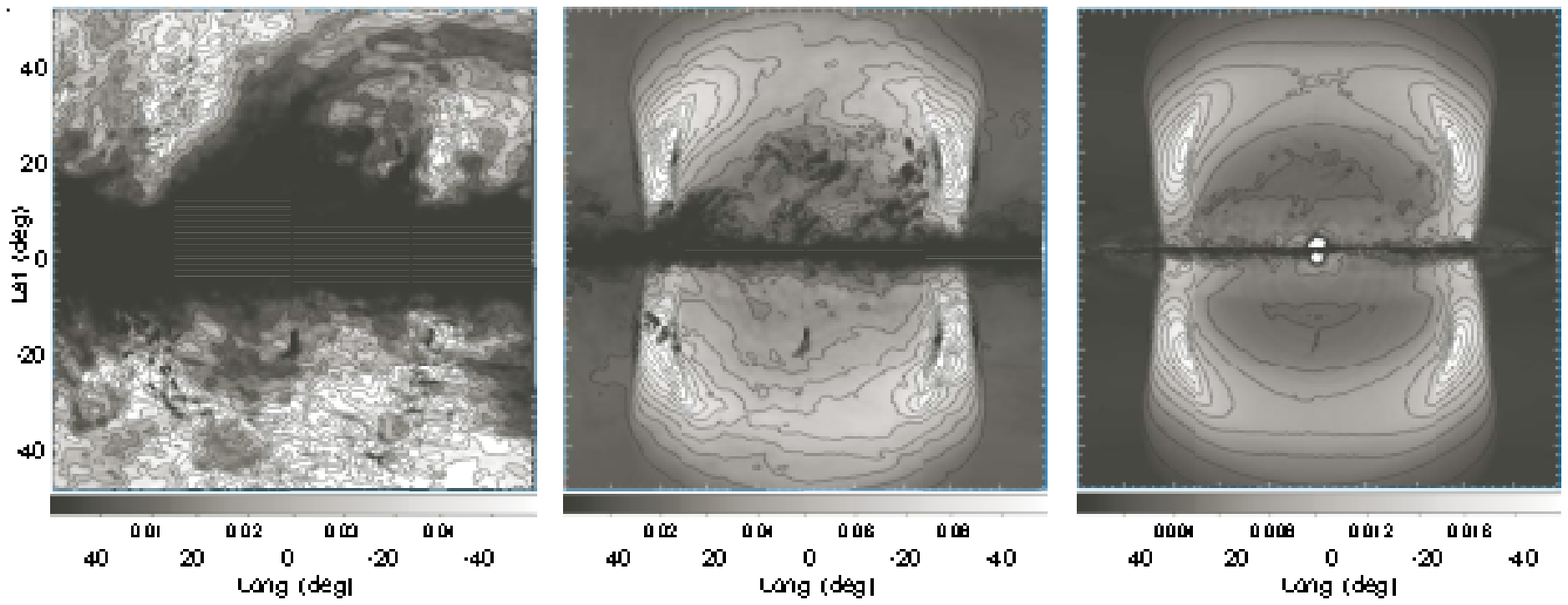} 
\end{center} 
\caption{(a) Simulated shadowed X-ray brightness as figure \ref{xsky1}a with contours at 0, 0.01, 0.02, ..., 0.16, 0.17 Jy str$^{-1}$. (b) Same as \ref{xsky1}b, but shadowed with the same contour levels. (c) Same as \ref{xsky1}c, but shadowed with the same contour levels. } 
\label{xsky2}  

\begin{center}
R2~~~~~~~~~~~~~~~~~~~~~~~~~~~~~~~~~~~~~~~~R4~~~~~~~~~~~~~~~~~~~~~~~~~~~~~~~~~~~~~~~~R7\\
\includegraphics[height=6.1cm]{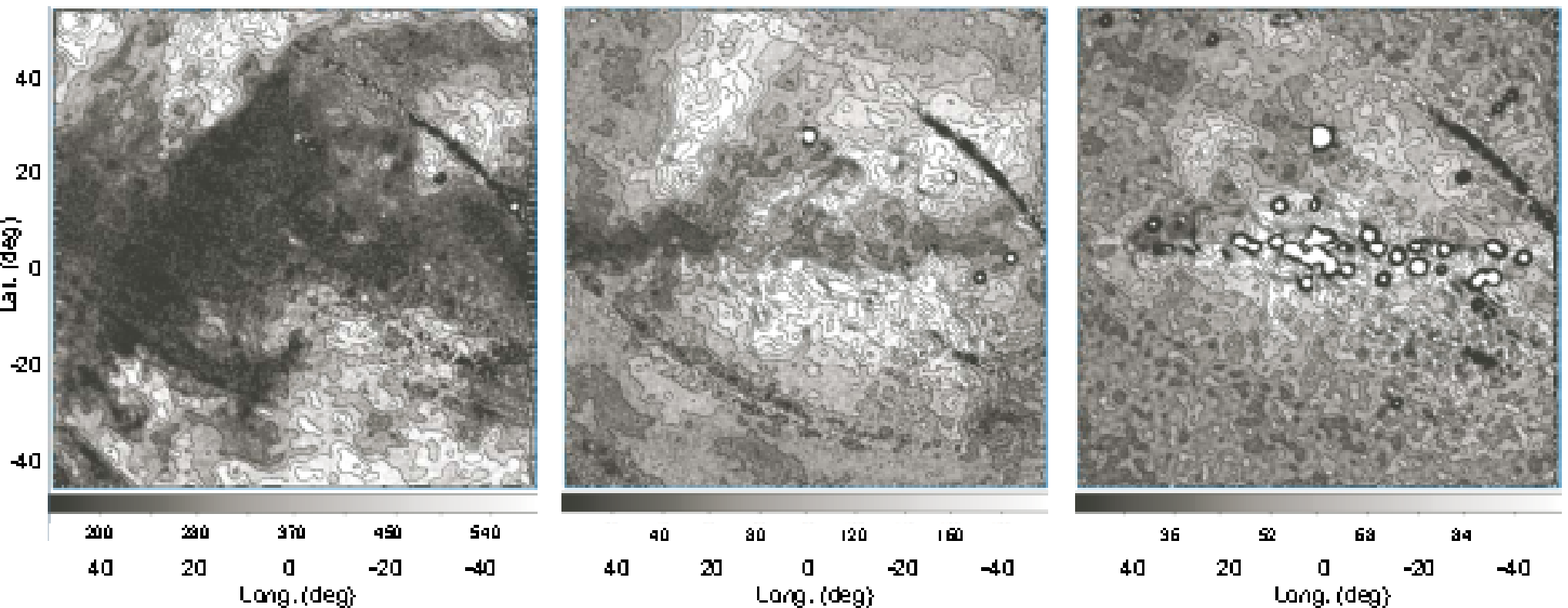}  
\end{center} 
\caption{ROSAT images in R2, 4 and 7 bands. Contour levels are: R2: 100, 200, 300, ...; R4: 30, 60, 90, ...; R7: 20, 30, 40, ... $\ROunit $.  
} 
\label{xsky2ro}  
\end{figure*}

\begin{figure*} 
\begin{center}
\includegraphics[width=12cm]{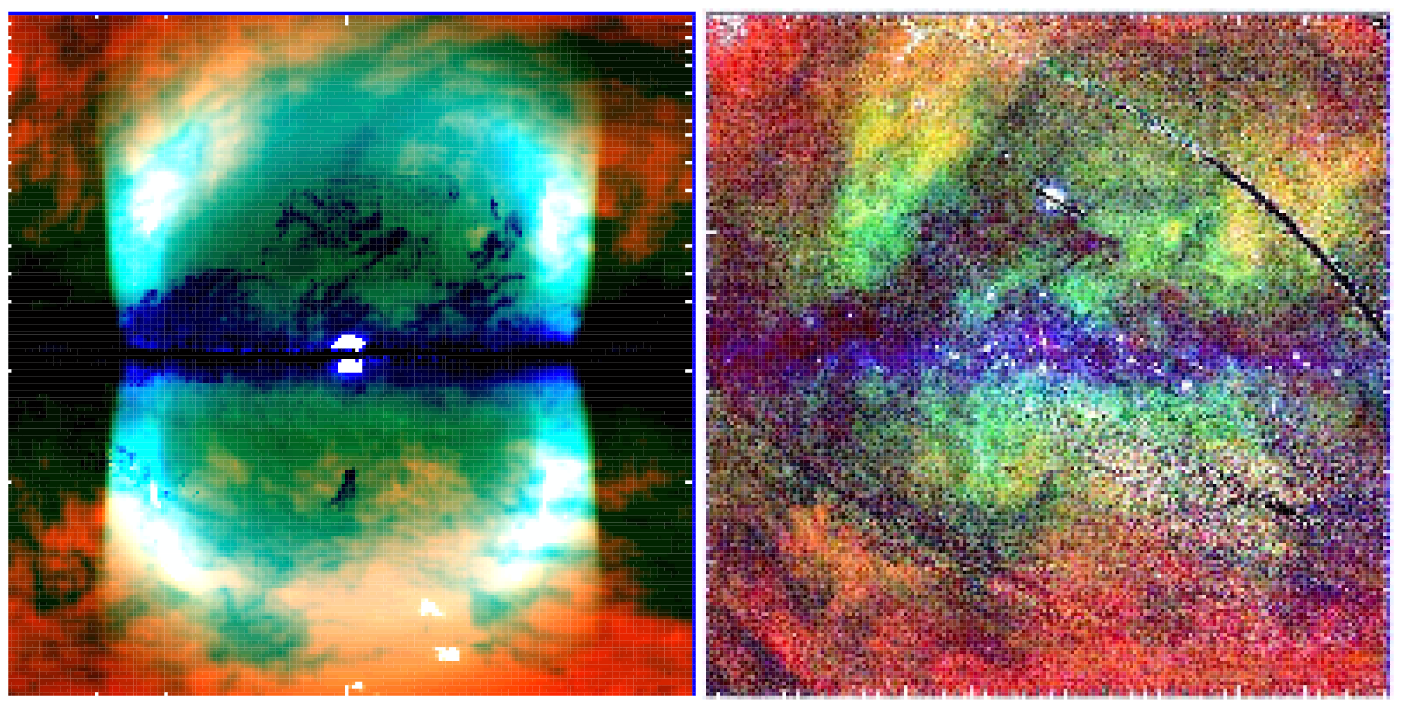}    
\end{center} 
\caption{Same as figures \ref{xsky2} and \ref{xsky2ro}, but in color coded representation by red for 0.25/R2, green for 0.7/R4 and blue for 1.5 keV/R7 bands (left: BHS, and right: ROSAT).}  
\label{xsky3} 
\end{figure*}

Figure \ref{xsky1} shows the thus computed distribution of the X-ray brightness on the sky in the central $\pm 50\Deg \times \pm 50\Deg$ region at $t=10$ Myr in for 0.25, 0.7 and 1.5 keV (approximating R2, R4 and R7 bands), showing the intrinsic intensity distributions without suffering from interstellar absorption. These maps are then multiplied by exp($-\tau_i$) to yield shadowed brightness distributions using $\tau_i$ distribution as shown in figure \ref{tau4}. Figure \ref{xsky2} shows the simulated shadowed X-ray sky compared with corresponding ROSAT images, both in contour representations.  The model brightness unit is Jy str$^{-1}$, while ROSAT brightness in $\ROunit$. Figure \ref{xsky3} is a color coded representation compared with the ROSAT color map. 
 
As in figure \ref{xsky1}, the BHS shows up as the dumbbell shaped structure of symmetric spurs extending toward the galactic poles. The observed North Polar Spur is well reproduced by the simulation. The western counter part of NPS (NPS-W) at $(l,b)\sim(-30\Deg,+10-40\Deg)$, the South Polar Spurs SPS-E and SPS-W are also seen in the ROSAT map, though not clear as in the north (see the next section for cross sections). 

The BHS spurs near the galactic plane are heavily obscured by the HI and molecular gas layers, spurs and clouds. Various shadow features are evident, silhouetting the BHS and halo X-ray emissions. Particularly, the R4 band NPS ridge is strongly shadowed by the Aquila Rift at $(l,b)\sim(25\Deg,12\Deg)$, where the Rift's ridge crosses the NPS nearly perpendicularly. The R7 band emission is less obscured, and shows more intrinsic structures. However, the observed ROSAT R7 band image is too noisy to show up the detailed structure of the spurs.

The 0.25 keV ($\sim$R2) emission is strongly absorbed by the local interstellar gas, and only high-latitude emission weakly remains as northern and southern polar caps. The simulated 0.25 keV brightness is an order of magnitude darker than the intrinsic brightness. This is due to heavy extinction with large optical depth at $|b|<\sim 30\Deg$, particularly, due to the HI and H$_2$ spurs and flares of the Aquila Rift. Although the simulated and ROSAT R2 band images are similar to each other, they may not be further compared seriously, because the observed R2 emission is a mixture of local emissions such as the Solar Wind Charge eXchange and the Local Hot Bubble.

\section{Discussion}

The shock front was shown to expand spherically in the initial $\sim 1$ Myr, and is elongated into the halos composing dumbbell shaped symmetric bubble features with respect to the galactic disk. Expected X-ray brightness distributions were calculated for the central $\pm 50\Deg \times \pm 50\Deg$ region at three different energies (0.25, 0.7, and 1.5 keV), approximately corresponding to R2, R4 and R7 ROSAT bands. Shadowing by the interstellar HI and H$_2$ gases was taken into account to compute the brightness distributions on the sky.

The simulated X-ray maps at elapsed time $t\sim 10$ My approximately reproduces the morphological features of the NPS and SPS as observed by ROSAT. The modeled emission measure was found to be greater than those from {\it Suzaku} observations at a couple of tens positions by a factor of $\sim 1.5$. Below, we consider about the implication of the results and discuss some problems exhibited by this study. 

\subsection{Simulation vs Observation}
\subsubsection{ROSAT count rate vs Jy$~str^{-1}$} 
The ROSAT data are represented in terms of count rate of the instrument in unit of counts s$^{-1}$ arcmin$^{-2}$, but not calibrated to intensity, whereas the simulated result is presented in unit of 
 ${\rm Jy~ str^{-1}}$
\footnote{${\rm 
 1~Jy= 10^{-26}w~m^{-2}~Hz^{-1}
 =2.41798\times 10^{-9} erg~cm^{-2}~s^{-1}~eV^{-1}
 }$}. 
In order to obtain quantitative comparison, we calibrated the ROSAT data into Jy str$^{-1}$ using the {\it Suzaku} observations of the emission measure ($EM$) and plasma temperature in the region near the Fermi Bubbles including NPS and southern spurs obtained by Kataoka et al. (2015).  

\begin{figure}
\begin{center}  
 \includegraphics[width=7cm]{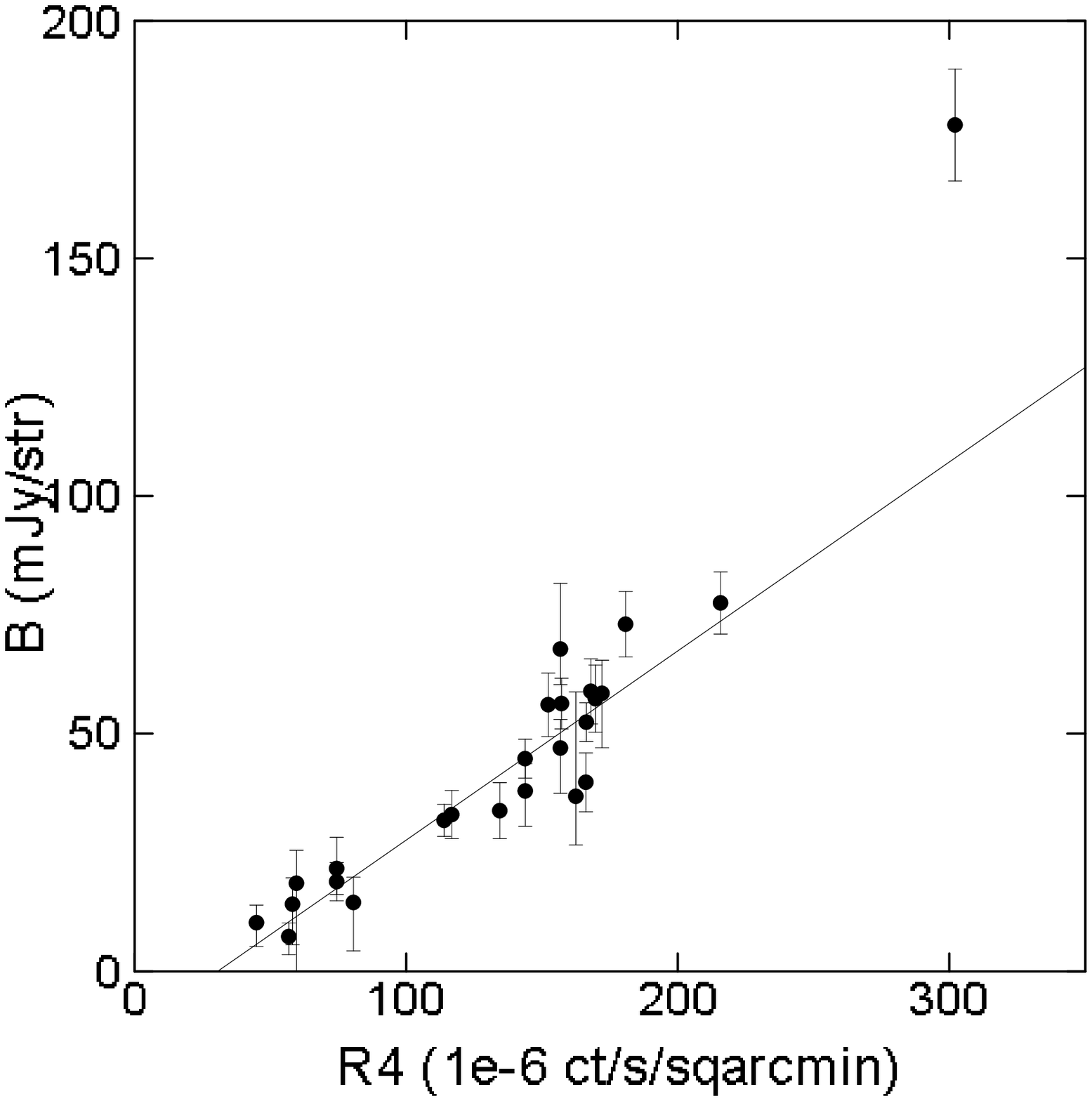}    
\end{center}
\caption{Surface brightness in mJy str$^{-1}$ plotted against R4-band count rate at positions where $EM$ and plasma temperature were determined by {\it Suzaku} observations (Kataoka et al. 2015).}
\label{calib} 
\end{figure}

X-ray brightness $B$ at photon energy $E=0.7$ keV was calculated by using the measured $EM$ for plasma temperature of $T=0.3$ keV (Kataoka et al. 2015) adopting the cooling function for a metallicity $Z=0.2\Zsun$ as shown in figure \ref{cfunc}. Individual values were also multiplied by exp$(-\tau)$ with $\tau$ being the optical depth for $N_{\rm H}$ at corresponding positions. Figure \ref{calib} shows the thus calculated brightness plotted against the ROSAT count rate in R4 band at the same positions.

Removing the exceptionally large value near the top-right corner, we applied the least-square linear fitting, and obtained the straight line. The slope of the fitting is adopted as the calibration factor $X({\rm R4})$ as 
$X({\rm R4})=0.398\pm 0.021 {\rm ~mJy ~str^{-1} (1.0\times 10^{-6} counts~s^{-1}~arcmin^{-2}})^{-1}$. 
Using the thus determined factor, we compare the intensities by simulation and observations.
 
\subsubsection{Intensities}
\begin{figure}
\begin{center} 
 \includegraphics[width=7cm]{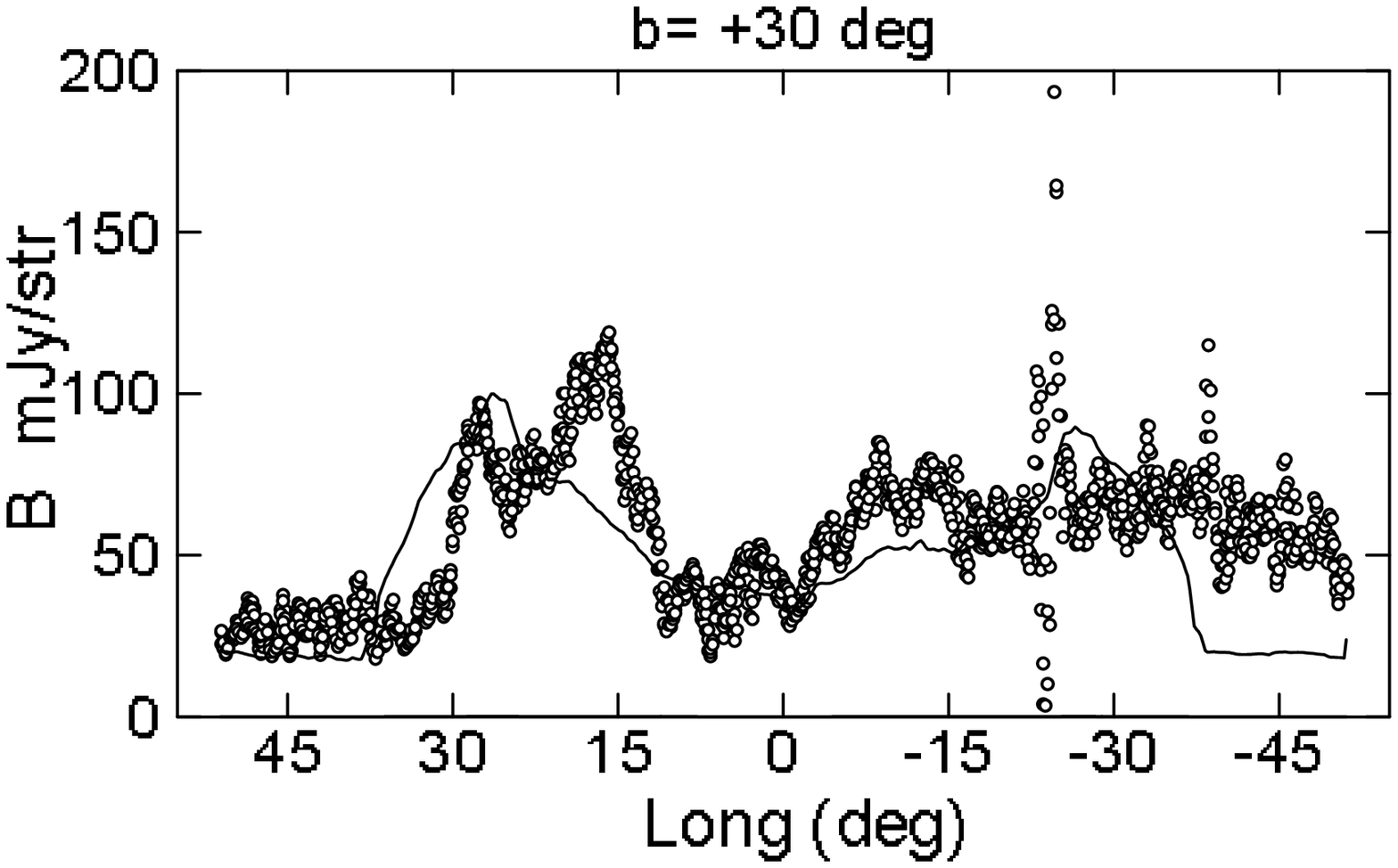}  \\ 
 \includegraphics[width=7cm]{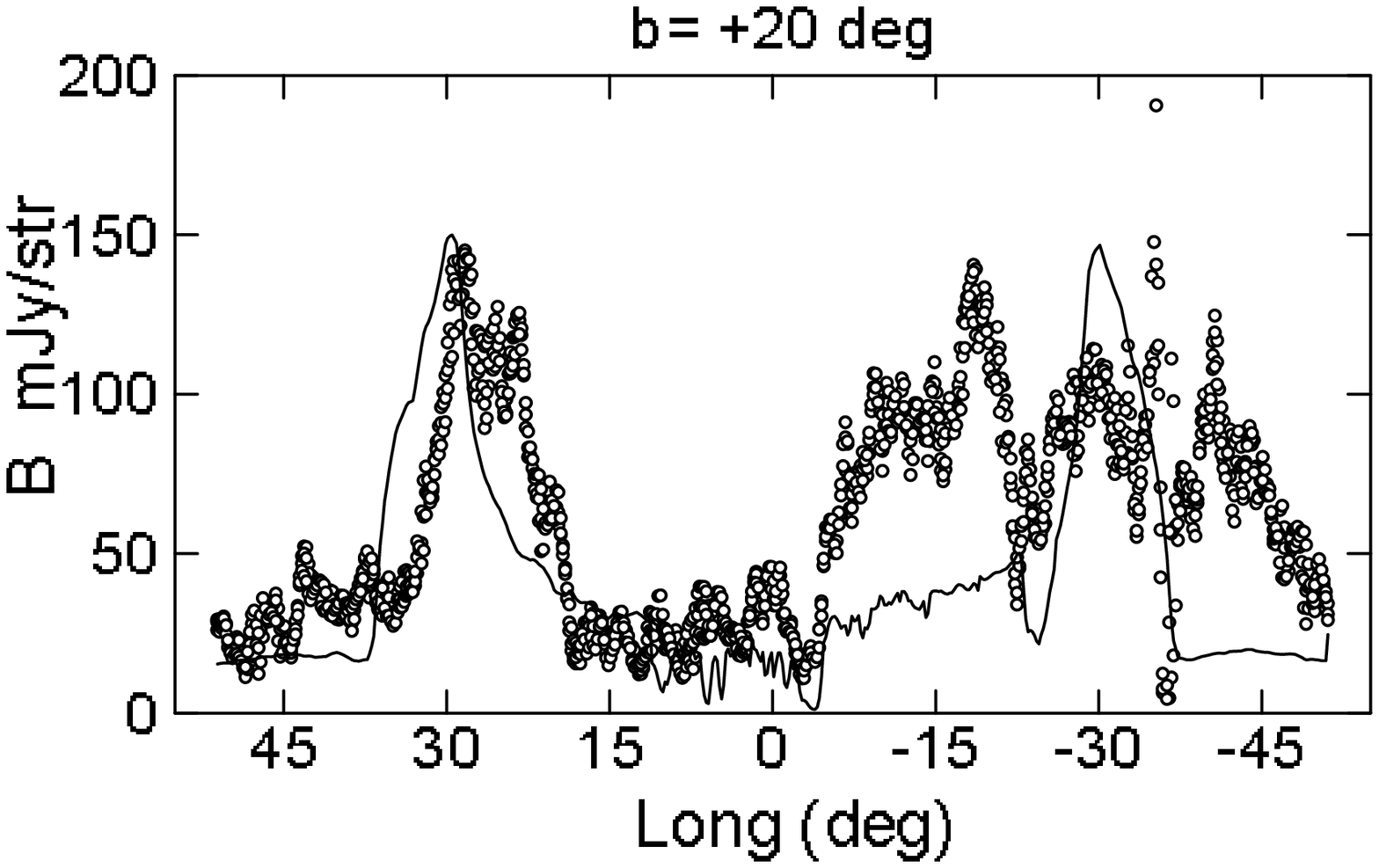}  \\ 
 \includegraphics[width=7cm]{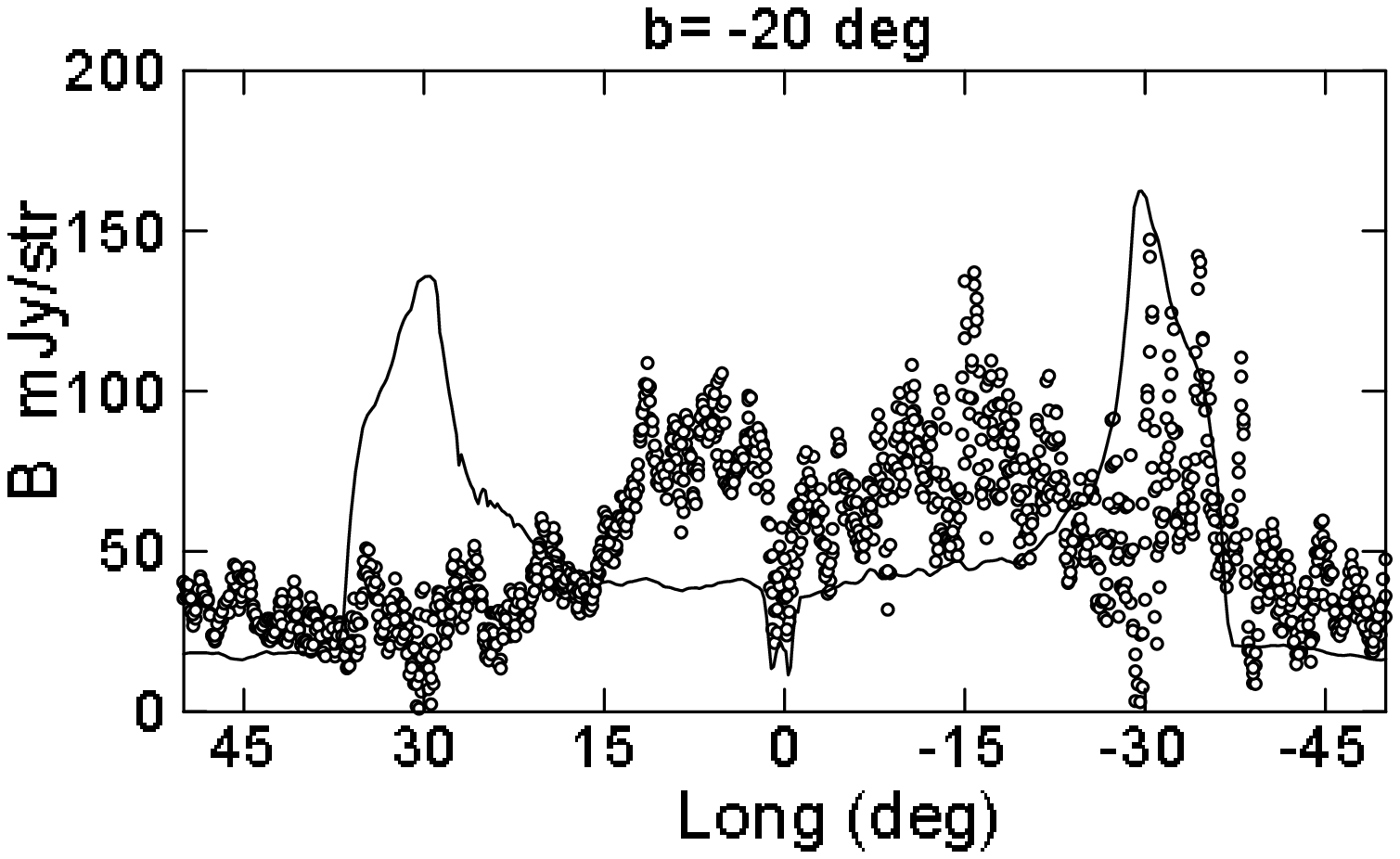}  \\ 
\includegraphics[width=7cm]{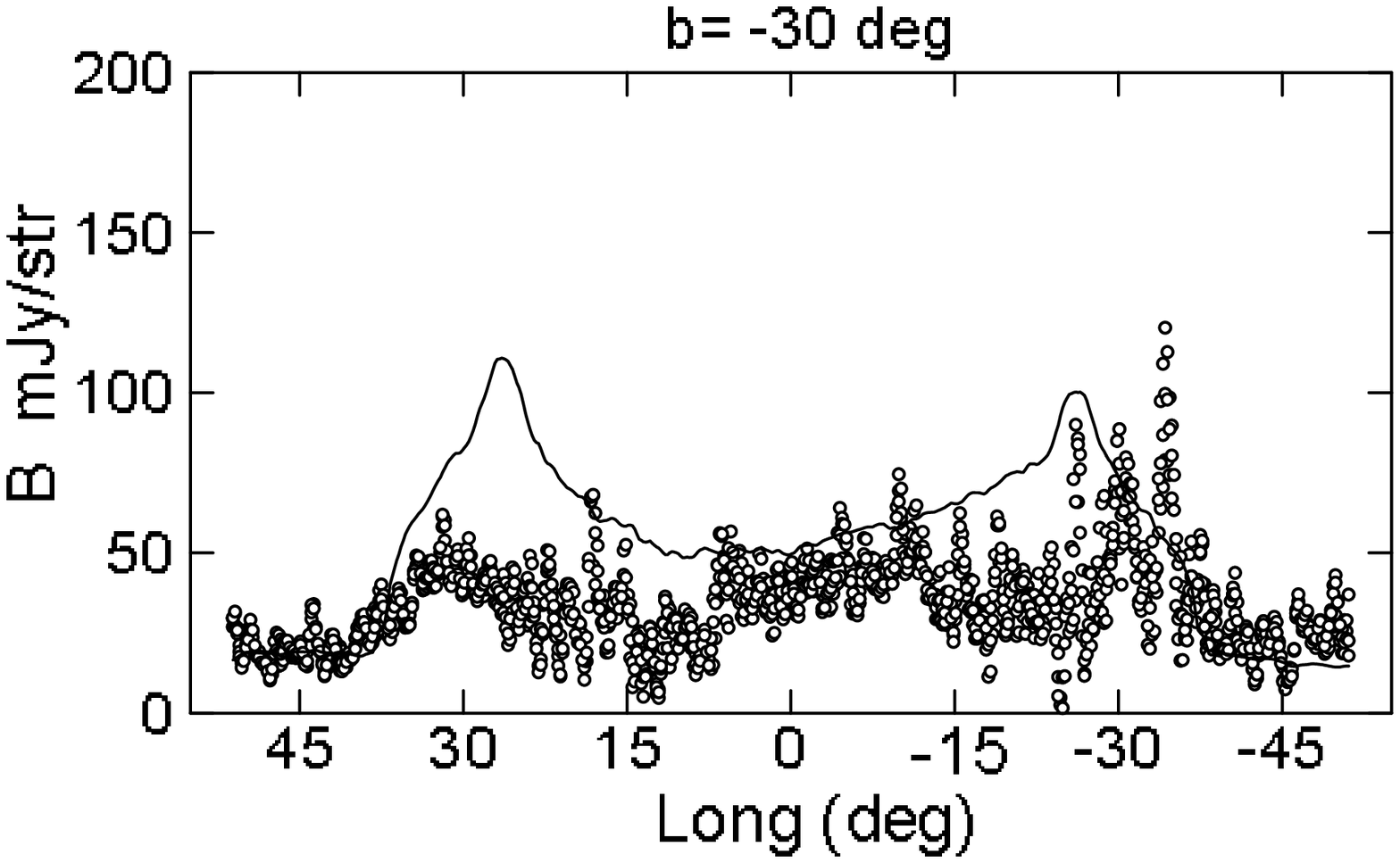}  
\end{center}
\caption{Simulated variations of 0.7 keV brightness along $b=+30\Deg,~ +20\Deg, ~-20\Deg$ and $b=-30\Deg$ (thick lines)  compared with those of ROSAT R4 band (circles). Both are shown by brightness in mJy str$^{-1}$.}
\label{cut} 
\end{figure}

Figure \ref{cut} shows cross sections of the R4 band intensity by observation (dots) and 0.7 keV by simulation (thick lines) along constant latitudes at $b=+30\Deg$ to $-30\Deg$. The plots were obtained as follows: the original ROSAT R4 map was convolved by a Gaussian function with a half width of 1 pixel ($0\Deg.2$). Then, pixel values along three rows at $b=+20\Deg \pm 0\Deg.2$ and $-30\Deg \pm 0\Deg.2$ were plotted by circles. The intensity scales are both in mJy str$^{-1}$.

The observed northern cross section at $b=+20\Deg$ exhibits double horn peaks of NPS and NPS-W. The NPS-W is contaminated by the bulge emission whose eastern half is heavily shadowed by the Aquila Rift. The simulated cross section shows the double horn feature symmetrically to the Galactic Center. 

The southern cross section at $b=-30\Deg$ by simulation, particularly SPS-E, well fits the observation in the shape. The SPS-W is again contaminated by the bulge emission as the round enhancement around the Galactic Center.

Although the agreement of peak intensities in the NPS is excellent, the simulated intensities are generally higher than the observations. We may eye-estimate the simulated-to-observed intensity ratios at the peaks in figure \ref{cut} to obtain $B_{\rm simu}/B_{\rm obs}=1.6\pm 0.8$. This ratio is directly related to the emission measure, and therefore, the gas density in the BHS and the ambient halo gas as $\rho_{\rm simu}/\rho_{\rm obs}\sim 1.3$.

\subsubsection{Metallicity}
Although we assumed $Z=\Zsun$ as the marginal value for the entire evolution of BHS, it is, in reality, variable with time and place. In the early phase of expansion, the BHS is dominated by the high-metal gas in the Galactic Center and the disk where $Z> Z_\odot$. As it expands into the halo, the BHS becomes dominated by the swept halo gas and intergalactic gas with $Z<\Zsun$.
However, the radiative cooling does not contribute much to the dynamical evolution of the BHS, namely simulations assuming a different metallicity would result in almost the same result.

On the other hand, the X-ray brightness was calculated for the observed metallicity in the galactic halo with $Z=0.2\zsun$.
Hence, the metallicity in the simulation and in the brightness calculation is not fully consistent with each other. A consistent modeling would be a subject for the future.

\subsection{Energetics}

\def\Ek{\E_{\rm kin} } 
\def\Eg{\E_{\rm grav} } 
\def\Ec{\E_{\rm rad} } 
\def\Et{\E_{\rm th} } 
\def\Mbhs{M_{\rm BHS} } \def\Mgal{M_{\rm gal} }
\def\ergs{{\rm ergs} }
 
We have presented the result of BHS simulation for a continuous energy injection (type C injection) with $\E=4\times 10^{56}$ ergs in the last 10 Myr presuming recurrent starbursts in the Galactic Center (SB model). The total injected energy is smaller than those estimated for super winds in starburst galaxies (Heckman et al. 1990), while greater than that for the starburst galaxy M82 (Lacki et al. 2014). It is also greater than that proposed for the Galactic Center in order to produce the Fermi Bubbles (Crocker et al. 2015). 

\subsubsection{Energy partitioning in the BHS}
The total injected energy $\E_0$ is finally transformed to the thermal $\Et$, kinetic $\Ek$, and gravitational $\Eg$ energies of the BHS as well as to the radiatively lost energy by cooling $\Ec$:
\begin{equation}
\E_0=\Et+\Ek+\Eg+\Ec.
\label{Esum}
\end{equation}
Individual energies are approximately estimated using the figures presented in section 2 as the following. The BHS mass is estimated to be $\Mbhs\sim 5 \times 10^7 \Msun$ for the both sides,  each with a radius $R\sim 4$ kpc and thickness $\Delta R\sim 1$ kpc and mean density of $n\sim 0.005~{\rm cm}^{-3}$.
The thermal energy is estimated for a mean temperature of $T\sim 3\times 10^7$ K, where  $\mathscr{R}$ being the gas constant. 
The kinetic energy is estimated for an expansion velocity $V_{\rm expa}\sim 400$ \kms. 
The gravitational energy can be estimated by $\Delta \Phi\sim G \Mgal(1/h_0 - 1/h)$, where $\Phi$ is the gravitational potential, $z_0\sim 2$ kpc and $z\sim 4$ kpc are the initial and the present heights of BHS gas. The galactic mass within a radius  $r \sim z_0$ kpc is estimated by $\Mgal \sim z_0 V_{\rm rot}^2/G \sim 1.8\times 10^{10}\Msun$ for a rotation velocity $V_{\rm rot}\simeq 200$ \kms. The radiative loss during $t\sim 10$ Myr is estimated for a cooling rate of $P(T)\sim 3 \times 10^{-23} \ergs ~{\rm cm^3~ s^{-1}}$ at $T\sim 3\times 10^7$ K and $n\sim 10^{-2}{\rm cm}^{-3}$. 

We now obtain:
\be
\Et=\int p dV \sim  \mathscr{R}  \Mbhs  T \sim 2.5 \times 10^{56}{\rm ergs},
\ee
\be
\Ek \sim 1/2 \Mbhs V_{\rm expa}^2\sim  8\times 10^{55} {\rm ergs},
\ee
\be
\Eg \sim \Mbhs \Delta \Phi \sim 2\times 10^{55}{\rm ergs},
\ee 
\be
\Ec\sim t \int  n^2 P(T) dV \sim 3\times 10^{54}\ergs,
\ee
and 
\be
\E_0=\Et+\Ek+\Eg+\Ec \sim 5\times 10^{56} \ergs.
\ee
The estimation using the figure may include an error of $\sim \pm 30$\%, that is the reason for the larger figure of total energy than the input intival value of $4\times 10^{56}$ ergs. 

The relative fractions of the energies compared to $\E_0$ are rather more accurate, and are: $\Et/\E_0\sim 0.7$, $\Ek/\E_0\sim 0.2$, $\Eg/\E_0\sim 0.06$, and $\Ec/\E_0\sim 0.01$. 
The small radiative loss implies that the BHS is almost adiabatic. This can be confirmed by the long cooling time, $t_{\rm cool}\sim 9\times 10^9$ yrs, far longer than the expansion time of the shell. This implies that the BHS dynamics is little depends on the cooling rate, and hence on the density and metallicity. Namely, the result will be not changed much even if we adopt smaller metallicity than the assumed value (1 solar). 

\subsubsection{Constraint on the total energy}
We thus conclude that the total energy on the order of $\E_0 \sim 4\times 10^{56}$ ergs is inevitably required in order to reproduce the observed X-ray NPS and associated spurs in the BHS model for the given halo density. Thereby, the emission measure (density), temperature (velocity), and the linear size of the NPS are the strongest observational constraint on the total energy.
 
If the halo density is decreased by a factor of 1/1.3 as indicated by comparison with the emission measure observed with {\it Suzaku}, the total energy will be eased by the same factor, considering the Sedov's similarity factor $(\E_0/\rho_0)^{1/5}$ in the shock wave. Thus, we conclude that the total injected energy to drive the NPS is $\E_0\sim 3\times 10^{56}$ ergs.

\subsubsection{SB model and star formation rate}
Given the total energy as above, the assumed rate of energy supply of $d\E/dt\sim 4 \times 10^{55}$ ergs Myr$^{-1}$ is required in the C-type injection model, though reduced by a factor of 1/1.3. If we consider an SB model, it requires a type II supernova rate as high as $\sim 4\times 10^5$ per 10 Myr, or 1 SN per 20 to 30 yrs in the Galactic Center, comparable to the SN rate in the entire Galaxy. The assumed inflow rate of $dM/dt \sim 1  \Msun$ yr$^{-1}$ is comparable, but slightly greater than the often quoted rate $dM/dt\sim 0.1-1\Msun$ yr$^{-1}$ (Kruijssen et al. 2013; Krumuholz and Kruijssen 2015). 
 Hence, it may be worth to consider other types of mechanism of energy injection energy in the Galactic Center. 

\subsubsection{AGN model with E-type energy injection}
An additional or alternative energy source to drive the BHS could be the E-type energy injection at the center. A powerful energy source without suffering from the problem about star formation and mass inflow would be the E-type explosion due to AGN activity at the Galactic nucleus, which occurs intermittently each with smaller energy (Totani et al. 2006). In this case the total energy is given by $\E_0=\E_{\rm SB}+\E_{\rm AGN}$, and the mass inflow rate is accordingly reduced with increasing AGN activity.

\subsection{Central structures}

\subsubsection{Mach cone} 

Figure \ref{machcone} shows a central $\pm 25\Deg \times \pm 25\Deg$ region in the simulated 1.5 keV brightness compared with ROSAT R7 image. In the simulation, a cone-shaped bow shock is recognized as bipolar inclined spurs emerging from the central region, which is a sonic boom, or Mach cone, produced by the interaction of the outflow with the dense wall of the holed disk. The Mach angle of the cone is $\mu\simeq 58\Deg$, indicating a Mach number $\mathscr{M}=1/{\rm sin~}\mu \sim 1.12$. This is consistent with the postulated outflow velocity and temperature of the gas in the corresponding region. 

The Mach cone morphologically reproduces the X-ray bipolar conical structure. However, the simulated cone is fainter compared to the observation. Brighter cone could be obtained, if the energy injection rate was not constant, but varied intermittently, so that the most recent injection was strong enough to enhance the cone's emissivity. This idea is consistent with the shorter time scale, $\sim 10^6$ yrs, suggested for the Fermi Bubbles with which the X-ray Claw feature in the southern spur is associated (Kataoka et al. 2013, 2015). 

\begin{figure*}
\begin{center} 
 \includegraphics[width=14cm]{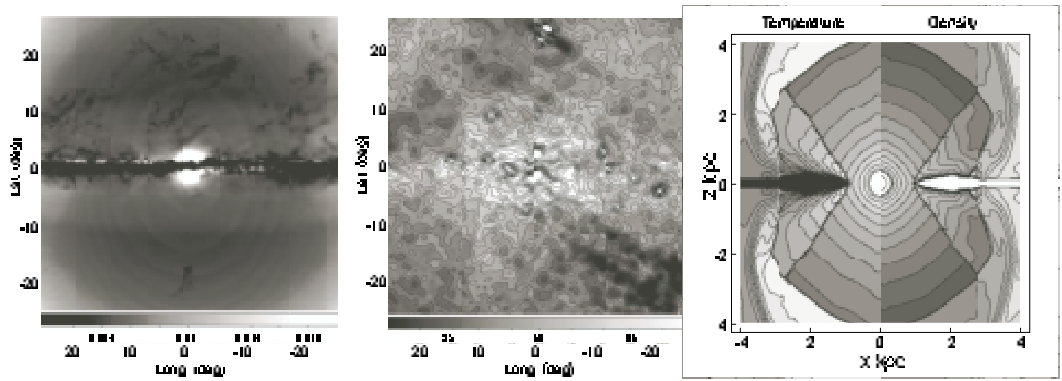}  
\end{center}
\label{machcone}
\caption{
The Mach cone by simulation at 1.5 keV (left), ROSAT R7 band (middle) in the central $\pm 25\Deg \times \pm 25\Deg$ region, and simulated temperature and density distribution in the $(x,z)$ plane for the central $\pm 4\times \pm 4$ kpc. The ROSAT image is point source removed, and the contours are at 10, 20, ... 200 ROSAT count rates. 
}
\label{machcone} 
\end{figure*}

\subsubsection{Central hot zone and expanding rings} 
\begin{figure}
\begin{center}  
 \includegraphics[width=8cm]{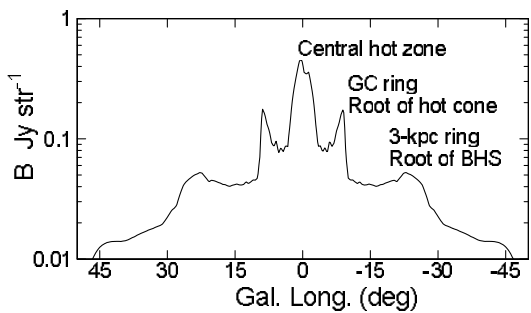}  
\end{center}
\caption{1.5 keV intrinsic brightness along the Galactic Plane, showing the central hot zone at GC, the 1 kpc ring at $l\sim 8\Deg$ corresponding to the root of hot cone, and the 3.3 kpc ring at $l\sim 24\Deg$.}
\label{cutGC} 
\end{figure}

Figure \ref{cutGC} shows a cross section of the simulated distribution of 1.5 keV brightness along the galactic plane. There appear three characteristic regions. A round hot zone of high-temperature plasma is produced by the injection of starburst thermal energy in the central $\sim 300$ pc at $|l|\le 3\Deg$. A shock-compressed ring of radius $\sim 1$ kpc at the root of the hot cone at $l=\pm 8\Deg$ appears as the two horn like enhancement of X-rays. A larger radius ring of a radius $\sim 3.3$ kpc appears as the broad maxima at $l\simeq 24\Deg$ corresponding to the root of the BHS.

These three structures may be related to  observed features in the Galactic Center at various wavelengths.  The central hot zone may correspond to the high-temperature plasma at $\sim 10$ keV observed in hard X-ray emissions (Uchiyama et al. 2013).   The 1 kpc ring may be a scaled up event of the various gas rings (Bania 1980; Sofue 1995; Law 2010;  Krumholz and Kruijssen 2015). The 3.3 kpc compressed ring associated with BHS coincides with the 3-kpc expanding ring observed in the HI line longitude-velocity diagram (Oort 1977). As readily known by a fast-mode MHD wave calculation (Sofue 1977), the compression occurs from outside of the disk by refraction and focusing of the shock wave propagating through the halo.

\subsection{BHS morphology: Asymmetry and intergalactic wind}    
In figures \ref{xsky3} and \ref{cut} the observed north-south and east-west asymmetry is evident in the sense that the NPS is a few times brighter than SPS and that the NPS-E and -W are inclined toward the west compared to the simulation. 

The north-south asymmetry may indicate that the electron density in the north is significantly higher than in the south. Such an asymmetry of the BHS could be produced by an intergalactic wind or a motion of the Galaxy through the intergalactic gas (Sofue 1994). The wind will yield higher density and pressure in the northern halo, whereas the southern halo suffers from rarefaction due to shading by the galactic disk from the wind. 

The axial asymmetry could also be explained by a wind from the east to west. Hence, we may speculate that an intergalactic wind blows from the galactic north-east toward south-west. Diagnosis of intergalactic wind using the BHS morphology would be a subject for future 3D simulations. \\

\noindent{\it Acknolwedgements}:  
A.H. is supported  by the Japanese Soc. Promotion of Sciences KAKENHI Grant No. 15K05014. Numerical simulations were in part carried out on computers at Center for Computational Astrophysics (CfCA), National Astronomical Observatory of
Japan. We thank the authors of the ROSAT All Sky X-Ray Survey, Columbia Galactic Plane CO-Line Survey, and the Bonn-Leiden-Argentine HI Line Survey for the archival data. \\

\noindent{\bf REFERENCES}\\
\def\r{\hangindent=1pc  \noindent} 

\r Ackermann M., et al., 2014, ApJ, 793, 64  

\r Anders E., Grevesse N., 1989, GeCoA, 53, 197  

\r Arnaud K.~A., 1996, ASPC, 101, 17 

\r Bania T.~M., 1980, ApJ, 242, 95 

\r  Bland-Hawthorn, J., and Cohen, M.\ 2003, ApJ, 582, 246

\r Carretti E., et al., 2013, Natur, 493, 66

\r Crocker R.~M., Bicknell G.~V., Taylor A.~M., Carretti E., 2015, ApJ, 808, 107  

\r Dame, T. M., Hartman, D., Thaddeus, P. 2001, ApJ 547, 792.   

\r Fang T., Jiang X., 2014, ApJ, 785, L24 

\r Foster, A.R., Ji, L., Smith, R.K., and Brickhouse, N.S.: 2012, ApJ 756, 128. 

\r Fujita, Y., Ohira, Y., \& Yamazaki, R.\ 2013, ApJ Letters, 775, L20.

\r Haslam, C.~G.~T., Salter, C.~J., Stoffel, H., and Wilson, W.~E.\ 1982, AA Suupl. 47, 1     

\r Heckman, T. M., Armus, L., Miley, G. K., 1990, ApJS 74,  833.
  
\r Inoue Y., Nakashima S., Tahara M., Kataoka J., Totani T., Fujita Y., Sofue 
Y., 2015, PASJ, 67, 56  

\r Kalberla, P.~M.~W., Burton, W.~B., Hartmann, D., et al.\ 2005, AA, 440, 775 

\r Kataoka J., et al., 2013, ApJ, 779, 57 
 
\r Kataoka J., Tahara M., Totani T., Sofue Y., Inoue Y., Nakashima S., Cheung C.~C., 2015, ApJ, 807, 77  
   
\r Kruijssen J.~M.~D., Longmore S.~N., Elmegreen B.~G., Murray N., Bally J., Testi L., Kennicutt R.~C., 2014, 
MNRAS, 440, 3370  

\r Krumholz M.~R., Kruijssen J.~M.~D., 2015, MNRAS, 453, 739 

\r Lacki B.~C., 2014, MNRAS, 444, L39 

\r Law C.~J., 2010, ApJ, 708, 474

\r  Miller M.~J., Bregman J.~N., 2013, ApJ, 770, 118 

\r  Miyamoto, M.; Nagai, R.	 1975PASJ 27, 533     

\r Mou G., Yuan F., Bu D., Sun M., Su M., 2014, ApJ, 790, 109

\r Najarro F., Figer D.~F., Hillier D.~J., 
Geballe T.~R., Kudritzki R.~P., 2009, ApJ, 691, 1816 

\r Nozawa T., Kozasa T., Habe A., 2006, ApJ, 648, 435

\r Oort J.~H., 1977, ARA\&A, 15, 295 

\r Sakai, K., Yao, Y., Mitsuda, K., et al.\ 2014, PASJ, 66, 83 

\r Sarkar, K.~C., Nath, B.~B., \& Sharma, P.\ 2015, MNRAS, 453, 3827 

\r  Raymond, J. C., Cox, D. P., Smith, B. W. 1976, ApJ 204, 290
  
\r Sakashita, S. 1971 ApSpSc 14, 431.

\r  Snowden, S. L., Egger, R., Freyberg, M. J., McCammon, D.,Plucinsky, P. P.,
  Sanders, W. T., Schmitt, J. H. M. M., Tr\"umpler, J., Voges, W. H.
1997 ApJ. 485, 125 

\r  Sofue, Y. 1977, AA 60, 327. 

\r  Sofue, Y. 1984,  PASJ  36,  539. 

\r Sofue, Y.  1995, PASJ 47, 527 

\r Sofue, Y. 2000, ApJ, 540, 224     

\r Su, M., Slatyer, T.~R., and Finkbeiner, D.~P.\ 2010, ApJ, 724, 1044    

\r Tahara M., et al., 2015, ApJ, 802, 91 

\r Totani T., 2006, PASJ, 58, 965 

\r Uchiyama H., Nobukawa M., Tsuru T.~G., Koyama K., 2013, PASJ, 65, 19 

\r van Albada, G. D.; van Leer, B.; Roberts, W. W., Jr. 1982 AA 108, 76

\r  Mair, G., Mueller, E., Hillebrandt, W., Arnold, C. N. 1988 AA 199, 114

\r Willingale R., Hands A.~D.~P., Warwick R.~S., Snowden S.~L., Burrows D.~N., 2003, MNRAS, 343, 995

\r Yao Y., Wang Q.~D., Hagihara T., Mitsuda K., McCammon D., Yamasaki N.~Y., 
2009, ApJ, 690, 143 

\end{document}